\documentclass[12pt,a4paper]{article}

\newif\ifblind
\ifdefined\blindbuild\blindtrue\fi

\usepackage[a4paper, margin=1in]{geometry}
\usepackage[T1]{fontenc}
\usepackage[utf8]{inputenc}
\usepackage{mathptmx}
\usepackage[scaled=0.92]{helvet}
\usepackage{amsmath}
\usepackage{graphicx}
\usepackage{booktabs}
\usepackage{threeparttable}
\usepackage{tabularx}
\usepackage{makecell}
\usepackage{longtable}
\usepackage[labelfont=bf, font=small]{caption}
\usepackage{microtype}
\usepackage{xcolor}
\usepackage{setspace}
\usepackage[round]{natbib}
\PassOptionsToPackage{hyphens}{url}
\usepackage[colorlinks=true, allcolors=blue!55!black]{hyperref}
\usepackage{flafter}
\newcommand{\datasetdoi}{\mbox{10.5281/zenodo.22177190}}

\newcommand{\Eone}{E1}
\newcommand{\Etwo}{E2}
\newcommand{\Ethreea}{E3a}
\newcommand{\Ethreeb}{E3b}

\newcommand{\mn}{\ensuremath{-}}

\newcommand{\aidisclosure}{The author conceived and designed the study, pre-registered it, assembled the data, wrote the first analysis code and the paper, and took every analytical and editorial decision. AI tools (Claude, Anthropic) assisted with code checking and debugging, reference checking, appendices, tables, figures and formatting. They also suggested grammar, style and formatting corrections in a limited capacity to text the author had written, each of which the author accepted or rejected individually. The author critically reviewed and verified all assisted output and takes full responsibility for the content.}

\title{The Price of Intelligence:\\A Quality-Adjusted Price Index for AI Services}
\ifblind
  \author{}
  \date{}
\else
  \author{Louis Yiven Zhu\thanks{Oxford Internet Institute, University of Oxford. Email \texttt{louis.zhu@oii.ox.ac.uk}. The author conducted this research independently and received no external funding for it. An earlier version was accepted at the NeurIPS 2026 Workshop on Economics for Machine Learning, and the author thanks its three anonymous reviewers. The author also thanks Professor Carl-Benedikt Frey, Professor Greg Taylor and Dr Thomas Robinson for their insights. The author further thanks Epoch AI, Artificial Analysis, the Internet Archive, OpenRouter and LMArena for maintaining the public records on which the panel rests.}}
  \date{November 2026}
  \hypersetup{pdftitle={The Price of Intelligence: A Quality-Adjusted Price Index for AI Services}, pdfauthor={Louis Yiven Zhu}}
\fi

\begin{document}
\maketitle

\begin{abstract}
\noindent How fast AI inference prices have fallen depends on how an index holds quality constant. We build quality-adjusted price indices for text and code inference sold through APIs, joining 21{,}024 posted prices to a capability scale fitted to 4{,}605 benchmark scores. From January 2025 to July 2026, matched-model methods record an annual fall of 0.07 in log price per token. Our pre-specified hedonic index falls at 0.51 a year in logs (95\% interval 0.11 to 0.89), entirely through the rising capability of the models on sale. Across three treatments of measurement error and two anchor sets it runs from 0.34 to 0.89. Excluding six anchors flagged for possible contamination leaves model rankings intact (rank correlation 0.998) yet lowers the rate by 0.40 to 0.51 under two treatments. Agreement on rankings therefore does not license claims about levels or increments. Per completed task, the data do not establish a quality-adjusted decline.

\end{abstract}

\section{Introduction}
\label{sec:intro}

\begin{figure}[htb]
  \centering
  \includegraphics[width=0.92\textwidth]{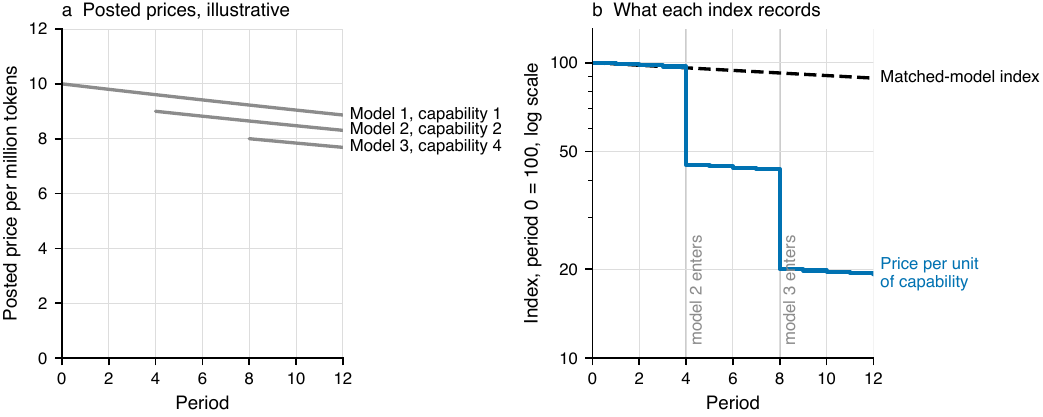}
  \caption{The estimand, drawn with cartoon data. (a) Three models enter at periods 0, 4 and 8 with capability 1, 2 and 4. Each enters at a slightly lower posted price than the last, and each price then drifts down slowly. (b) A matched-model index follows each model's own price and records only that drift. The price per unit of capability falls in steps at each entry, where each entrant sells more capability at a similar price. The hedonic index of Section~\ref{sec:indices} estimates the second path from data. The values are illustrative and carry no estimate.}
  \label{fig:estimand}
\end{figure}

Price indices for AI inference disagree about how fast prices fall, and the disagreement comes from how each index holds quality constant. A matched-model index follows each model's own posted price from month to month, and a model enters the comparison only in its second priced month. A hedonic index regresses posted prices on a measure of quality and reads the price change at constant quality from the regression's month effects. The difference matters in this market because quality arrives through new models, each of which enters at a price close to those of the models it surpasses. A matched-model index therefore records little of the fall in the price of capability, as Figure~\ref{fig:estimand} illustrates. In this paper we measure that fall with a hedonic index whose quality measure is a capability scale estimated from public benchmark scores.

Our main finding is that quality-adjusted prices per token fell about seven times as fast as matched-model methods record. Over January 2025 to July 2026, matched-model methods record a fall of 0.07 a year in the log of the posted price per token. Our pre-specified hedonic index falls at an annual log rate of 0.51 (95\% interval 0.11 to 0.89), a fall of about 40\% a year. Over the same months the average posted price of the benchmarked models on sale rises, by 0.07 a year in logs. The whole of the quality-adjusted decline therefore comes from the rising capability of the models on sale. We stress that the index describes the schedule of posted list prices against capability. It measures neither the prices that buyers paid after discounts nor the welfare they gained.

The panel behind these estimates covers text and code inference sold through application programming interfaces (APIs), on the capabilities that public benchmarks measure. It joins 21{,}024 posted prices for 3{,}208 models from 86 providers to a capability scale fitted to 4{,}605 benchmark scores for 782 models, and we recover the price history month by month from archived pricing pages. Image, audio and video generation lie outside the panel. On these data we ask three questions. \textbf{RQ1} asks how fast the quality-adjusted price per token fell from January 2025 to July 2026, and how much of that fall matched-model methods record. \textbf{RQ2} asks whether the fall survives the treatment of measurement error in capability and a validity audit of the capability scale. \textbf{RQ3} asks whether the quality-adjusted price per completed task fell over the months in which we can measure it.

Because capability enters the index as an estimate, the rate depends on how we treat the error in that estimate. Our pre-specified treatment imputes plausible values of capability from the benchmark scores alone. An imputation of this kind omits price and therefore attenuates the price gradient on capability, here to 0.61 of its value at the point estimate \citep{sterne2009multipleImputation}. The pre-specified estimator is thus the most conservative cell among those that use every anchor. Regression calibration, one of the most popular corrections for error in a covariate \citep{keogh2020measurementError}, gives 0.63 to 0.79 a year across its specifications. Across three treatments of measurement error and two sets of anchor benchmarks, the rate runs from 0.34 to 0.89, and four of the six intervals exclude zero. On any of these estimates matched-model methods miss most of the decline, 86\% on the pre-specified estimator (37\% to 97\%) and 78\% to 92\% across the six cells.

The decline also arrives in steps, in the months when new models enter. Two months, February and September 2025, carry 50\% of the cumulative change in the pre-specified index. In each of the three largest monthly movements, the listings priced in both months move the log index by at most 0.002 in either direction. The movement comes with the entrants, among them DeepSeek-R1 and o3-mini in February, GPT-4.1 mini and nano in May and GPT-5 mini and nano in September 2025. From January to July 2026 the index does not fall at all. Its seven movements sum to $+0.05$, with an interval from $-0.19$ to $+0.28$ that includes zero. The timing explains why matched-model methods miss so much, since they compare only the prices of continuing listings and those prices barely moved in the months that carry the decline.

Our second contribution is a pre-registered validity audit of the capability scale, organised as a chain of four inferences from benchmark scores to the index \citep{kane2013validity}. The scoring inference fails. Six of the nine anchors that the registered secondary test can reach show differential functioning across release cohorts, a signature consistent with training-data contamination. Seven of the ten anchors predate nearly every model in the panel and so admit no comparison under the registered primary test. What the audit shows about the index matters more than the flags themselves. Excluding the six flagged anchors leaves model rankings intact at a rank correlation of 0.998. Under regression calibration, however, the exclusion lowers the annual log rate by 0.40 (0.23 to 0.59), and at the point estimate it lowers the rate by 0.51 (0.30 to 0.74). Plausible values damp the shift to 0.17 ($-0.32$ to 0.67). The mechanism is that the exclusion leaves the price term unchanged and cuts the capability term to 52\% and 47\% of its value with every anchor under those two treatments. A hedonic index consumes the spacing of the capability scale, and agreement on rankings does not license claims about levels or increments.

For the price per completed task, our third question, the data do not establish a quality-adjusted decline. All twelve rate and wedge intervals of the three per-task series span zero. The token counts behind those series come from a single measurement by Artificial Analysis in July or August 2026, carried back to every earlier month, and the series cover only 14 of the 87 models on the cost frontier in the data that \citet{emberson2026plungingPrice} release. They cannot test the decline that those authors report.

These results extend a literature on quality change in fast-improving products. \citet{byrne2018semiconductorPrices} measure microprocessor quality with benchmark performance scores and find that a matched-model producer price index understated the hedonic decline once the posted prices of existing chips became sticky over their life cycles. AI inference shows the same mechanism. Entrants carry the quality change, and the prices of continuing models move little. The difference is that our quality characteristic is itself an estimate from benchmark scores of uneven provenance, and those benchmarks saturate and may leak into training data. We therefore carry the estimate's error into the index and audit the spacing of the scale on which the index rests. The question has practical weight. The Bureau of Labor Statistics prices software in the consumer price index by matched models without hedonic adjustment \citep{barbarino2026softwareInflation}, and \citet{makridis2026countingAI} propose experimental measures that use benchmark scores to adjust software price indices. Our audit asks what such an adjustment rests on.

The paper proceeds as follows. Section~\ref{sec:lit} places our work in three literatures. Section~\ref{sec:data} describes the data, and Section~\ref{sec:capability} the capability scale with its three treatments of measurement error. Section~\ref{sec:indices} defines the indices and the headline window. Sections~\ref{sec:results}, \ref{sec:audit} and~\ref{sec:pertask} report the results per token, the validity audit and the bounded per-task result. Section~\ref{sec:limits} states the limitations in decreasing order of consequence, and Section~\ref{sec:conclusion} concludes.

\section{Related literature}
\label{sec:lit}

Our work draws on three literatures. The first is the hedonic measurement of fast-improving products, which is settled practice with well-documented stakes for measured output. Hedonic price indices for computers set the template that later work on information technology followed \citep{berndt1993computers, triplett2006hedonic}, and mismeasured deflators for information technology carry directly into measured real output \citep{byrne2016mismeasurement}. \citet{byrne2018semiconductorPrices} price microprocessors on benchmark performance scores and find that the matched-model producer price index understated the decline once the posted prices of existing chips became sticky. Cloud computing posed the same problem for a service. \citet{byrne2021cloudComputing} build hedonic price indices for the compute and database services of Amazon Web Services and a matched-model index for its storage service. \citet{coyle2018cloudAccounting} divide Amazon's prices by a unit of performance, the EC2 compute unit, for want of observations that could support a hedonic regression. At the Bureau of Labor Statistics, \citet{sawyer2023ppiCloud} estimate time-dummy hedonic models for the producer price index of cloud services. Their quality variable is either a provider's own performance benchmark or the characteristics of the microprocessors, and they present the models as an exploration that the bureau might implement in future.

The index-number literature supplies our estimator and its alternatives. The rolling-window time-dummy method with splicing gives our headline index \citep{dehaan2014rollingWindow}, and \citet{diewert2009imputation} set out its trade-off against hedonic imputation. \citet{ehrlich2026qualityAdjustment} compare hedonic indices with exact demand-based indices on item-level transaction data at scale. \citet{bajari2025hedonicAI} use machine learning to generate quality characteristics from the text and images that describe products. Hedonic theory licenses one interpretive boundary. The index recovers an equilibrium price schedule \citep{rosen1974hedonic}, and \citet{pakes2003hedonic} shows that it bounds the true cost-of-living change from above without measuring it. In all of this work the researchers observe the quality characteristic or generate it from observed product data. None of it prices a product whose quality characteristic is itself an estimate from test scores that saturate and may leak into the training data of the products they score.

The second literature tracks the price of AI inference directly, and most of it follows the cheapest model that reaches a fixed level of benchmark performance. \citet{appenzeller2024llmflation} finds that the cost of a model of equivalent performance falls about tenfold a year. \citet{cottier2025epochPrices} find falls of between nine and nine hundred times a year depending on the threshold, and \citet{gundlach2025priceOfProgress} find five to ten times a year for a given level of benchmark performance. \citet{emberson2026plungingPrice} extend the threshold method to the cost of completing tasks across models, reasoning settings and token budgets, and they estimate a fall of about thirteen times a year since 2023. These studies measure an efficiency frontier. A deflator must do more, since it has to aggregate across the whole quality spectrum and carry the uncertainty of its quality measure. Their rates also come from other periods and thresholds. The nearest comparable object in our panel is its lower envelope, which falls more slowly (Section~\ref{sec:results}).

Two studies come closer to a price index. The OECD builds quality-adjusted indicators of AI model prices from model characteristics, prices and providers, and its aggregate across modalities falls about 80\% between March 2023 and January 2025 \citep{andre2025oecdPrices}. \citet{demirer2025emergingMarket} document the structure of the market from the usage data of OpenRouter and Microsoft Azure, including wide price dispersion among models of comparable capability, and \citet{demirer2026jepMarket} describe how firms buy and sell AI. Epoch's Capabilities Index equates benchmark scores into one capability measure with bootstrap intervals \citep{epoch2026eci}, and our audit uses it as an external criterion. To this work we add three things. Our index carries the measurement error of capability, we compare it with the matched-model methods that agencies use, and we audit the scale on which it rests.

The third literature supplies the audit and its targets. Psychometrics treats validity as a property of the interpretation and use of scores \citep{messick1989validity} and organises its evaluation as an argument \citep{kane2013validity}. It also supplies equating and linking \citep{lord1980irt, vanderlinden2016handbook}, plausible values for statistics built on estimated abilities \citep{mislevy1991plausibleValues} and tests of differential item functioning \citep{holland1988mantelHaenszel}. From epidemiology we take the treatments of missing data and of measurement error in a covariate that we compare \citep{sterne2009multipleImputation, keogh2020measurementError}. The evaluation literature documents contamination \citep{sainz2023contamination}, devises statistical tests that detect it \citep{oren2023provingContamination} and records saturation \citep{ho2025rosettaStone}. Item response theory has served that literature before, for comparing test sets and for selecting efficient benchmarks \citep{vania2021comparingTestSets, rodriguez2021evaluationExamples, polo2024tinyBenchmarks}. Our audit asks a question that this literature has not put to a price statistic. A hedonic index consumes the spacing of a capability scale as well as its order, and we ask whether that spacing survives the removal of suspect benchmarks.

Finally, the battery contains two instruments built expressly to measure economically valued work, GDPval \citep{gdpval2025} and the Remote Labor Index \citep{rli2026}. They enter the capability estimate on only eleven and twelve models respectively and anchor nothing, so their influence on the index is small. The occupational suite of \citet{wan2026economicevals} and the agentic benchmarks of \citet{fish2026econevals} are absent from the corpus we consume. \citet{epochEconReview2026} review three other benchmarks of the class, namely the Remote Labor Index, GDPval and APEX-Agents. \citet{espeland1998commensuration} study commensuration, the reduction of different qualities to one metric, and our audit measures one cost of such a reduction for a price statistic.

\section{Data}
\label{sec:data}

We build the panel from six public sources, each with a distinct role. Epoch AI's Benchmarking Hub supplies benchmark scores, model metadata and release dates \citep{epoch2026benchmarkingHub}. Artificial Analysis supplies posted prices and per-task token counts through its current pages \citep{artificialanalysis2026}, and the Internet Archive preserves the history of those pages \citep{internetarchive2026}. Archived pages embed the site's application state as structured data even where the visible table renders only in the browser, and parsing that state recovers a monthly price history that no rendered table carries. OpenRouter supplies prices for cross-checking, archived usage shares and most of the August 2026 live pull \citep{openrouter2026}. LMArena supplies a criterion of human preference for the audit \citep{lmarena2026}, and provider pricing pages supply an independent check on individual prices.

The resulting price panel holds 21{,}024 observed price rows for 3{,}208 models from 86 providers over 31 months, from February 2024 to August 2026. We record a further 601 explained gaps inside that span and never interpolate them. The score table holds 4{,}605 scored cells on 3{,}796 distinct pairs of model and benchmark across 64 benchmarks. Of those pairs, 357 carry more than one published score and contribute 809 additional rows, and the estimator treats each score as its own observation. The capability model estimates $\theta$ for the 782 models that these cells reach. The hedonic join then holds 6{,}158 priced observations on 701 endpoints for 500 of those models, where an endpoint is one model's listing on one provider's channel. These denominators differ, and every count below names the one it uses.

Archived captures price every month of the headline window, from January 2025 to July 2026, in 18{,}429 rows. August 2026 is the only month that we price from a live pull, and it adds 574 rows, 395 from the OpenRouter models API and 179 from Artificial Analysis model pages. The live pull reaches listings that the archive never captured. August 2026 is the only month in which 462 of the 3{,}208 priced models appear, and it is the first priced month for 141 of the 496 models that both of our capability fits price. Section~\ref{sec:indices} explains why the headline window ends in July 2026 for this reason.

Score provenance varies, and it varies with benchmark age. Of the scored cells, 23.3\% come from Epoch's own evaluations, 33.9\% from third-party leaderboards, 26.0\% from the model's developer and 16.8\% from unattributed sources. Developer-reported scores concentrate on the ten anchors that link the generations of the capability scale, where they supply 40\% of the 1{,}577 cells against 19\% of the 3{,}028 cells on the other benchmarks. Provenance and benchmark age therefore move together over the stretch where the equating carries most weight, and the audit has to confront that overlap. Appendix~\ref{app:carried} reports the provenance of each benchmark's scores.

Every number in the panel is traceable. Each price row carries the address and capture time of the page we read it from, and every analysis file records a hash of each of its inputs. We hand-checked 40 price rows against their recorded sources and verified all 40 without correction. We drew them from the most recent month and from the two current-page sources, so none of the 40 comes from the archive. For the archive we release a generated 5\% sample of 1{,}051 rows for inspection, 1{,}022 of them archival, and we have not checked that sample by hand.

Every price is a posted list price on one named first-party channel per provider, blended at three parts input to one part output per million tokens. Prompt caching, batch tiers, committed-use discounts and enterprise agreements can push transaction prices below list. Enterprise pricing is largely proprietary \citep{soto2026aiBuildout}, and the public record does not show these discounts. If they grew over the window, transaction prices fell faster than the list prices we measure, and we cannot observe that change. List prices also differ by channel. One model in our sample carried three simultaneous prices at its release, namely its developer's discounted launch tier, the aggregator's record of the first-party price and the prices of resellers on a routing marketplace. For that reason we fix one channel per provider by rule.

The per-task price multiplies a model's posted price by the output tokens it uses to complete a fixed evaluation suite. Artificial Analysis measured those token counts in July or August 2026, and we carry each model's count back to every earlier month. As a result, 66\% of the per-task observations carry dates before July 2026 and rest on a count measured later (Section~\ref{sec:pertask}). The scope of the panel follows from its sources. It covers text and code inference sold through APIs, on the capabilities that the benchmarks measure, and image, audio and video generation lie outside it.

\section{Measuring capability}
\label{sec:capability}

No single benchmark can serve as the quality characteristic of this market. No benchmark scores every model in the panel, benchmarks saturate and retire within the sample, and any one benchmark would carry its own defects into the index. We therefore estimate one capability number per model, $\theta_m$, from all scored cells jointly. The estimator is a latent-variable model from educational measurement, and it serves here as an aggregator of characteristics \citep{lord1980irt, vanderlinden2016handbook}. The computer price literature built its quality ladder from characteristics such as clock speed, memory and hard-disk storage \citep{berndt1993computers}. Ours comes from the pattern of scores across benchmarks.

We fit a logit-scale form of Samejima's continuous response model to each score \citep{samejima1973continuous}. Writing $s_{mk}$ for the score of model $m$ on benchmark $k$ as a proportion, the model is
\begin{equation}
\operatorname{logit}(s_{mk}) = a_k\,(\theta_m - d_k) + \epsilon_{mk}, \qquad \epsilon_{mk} \sim \mathcal{N}(0, \sigma^2),
\label{eq:crm}
\end{equation}
where $d_k$ is the difficulty of benchmark $k$ and $a_k$ its discrimination. A model scores higher on easier benchmarks and higher the more capable it is, and discrimination records how sharply a benchmark separates strong models from weak ones. We estimate one residual standard deviation $\sigma$ for all benchmarks, at 0.449. The data force that restriction on two counts. A beta likelihood on proportions has no maximum on this table, and per-benchmark residual variances collapse on thinly scored benchmarks (Appendix~\ref{app:carried}). Scores of exactly zero or one move inward by $\varepsilon = 0.001$ before the logit.

Identification and equating fix the scale and carry it across benchmark generations. We pin GPQA Diamond at $a = 1$ and $d = 0$ and link benchmark generations by two-stage fixed-parameter equating. The first stage fits the anchor benchmarks alone. The second holds their parameters fixed and estimates every other parameter against them. Fixing the anchors keeps thin recent benchmarks from bending the scale that links the eras, and it makes that link explicit enough for the audit to test. The cost is that defects in the anchors propagate by construction. For that reason the audit tests the anchors first, and its contamination sensitivity refits the scale on a new anchor set.

A rule fixed before estimation selects the anchors. A benchmark qualifies when at least ten scored cells fall on each side of 1 July 2024 and its scores among models released after that date have a standard deviation of at least 0.10. The first condition ensures that a benchmark links the two eras, and the second excludes benchmarks that no longer separate recent models. Ten benchmarks qualify, and counting models gives the same list as counting cells. Rankings agree across them, with a median Spearman correlation of 0.809 over anchor pairs and 0.823 over pairs that share at least twenty models.

The scale has an arbitrary origin and unit, and neither affects the index. The month effects of a hedonic regression are invariant to any rescaling of $\theta$ that stays constant over time. Drift in the equating over time would change them, and the audit's test of differential functioning exists to detect such drift.

An index built on an estimated characteristic inherits its measurement error, and here that error is far from uniform. The posterior standard error of $\theta$ has a median of 0.135 and runs from 0.033 to 3.337 across the 782 models. We compare three treatments of that error throughout the paper. The first, pre-specified in our registration, draws $L = 20$ plausible values of $\theta$ per model from the estimation posterior \citep{mislevy1991plausibleValues}. We compute every index once per draw and combine the draws by Rubin's rules \citep{rubin1987multiple}. We generate the draws from the benchmark scores alone, so they carry no information about prices. An imputation that omits the outcome of the analysis biases the estimated association toward zero \citep{sterne2009multipleImputation}, and the plausible values accordingly attenuate the price gradient on capability. Across the 18 windows of the headline estimator, the median ratio of the coefficient under the draws to the coefficient at the point estimate is 0.61.

The second treatment uses each point estimate $\hat\theta_m$ as if it were known. The third is regression calibration, one of the most popular methods for error in a covariate \citep{keogh2020measurementError}. It replaces each point estimate with its expected true value given the estimate,
\begin{equation}
\tilde\theta_m = \mu_m + \lambda_m\,(\hat\theta_m - \mu_m), \qquad \lambda_m = \frac{\tau^2}{\tau^2 + s_m^2},
\label{eq:rc}
\end{equation}
where $s_m$ is the model's standard error and $\tau^2$ the variance of true capability around the population mean $\mu_m$. Our primary specification sets $\mu_m$ to the mean capability of the models first priced in the same month. Two variants replace it with a linear trend in the first priced month or with one global mean. We estimate $\tau^2$ by the method of moments as the variance of $\hat\theta_m - \mu_m$ less the mean squared standard error. It takes 0.275, 0.376 and 0.825 under the three population models with every anchor. Calibration shrinks the least precise estimates most, toward the capability typical of their entry cohort. We calibrate once before resampling, so the intervals of the calibration cells carry price sampling alone.

We do not know which of the three treatments is correct on these data. Plausible values bias the gradient toward zero for the reason just given, and point estimates ignore the error altogether. Regression calibration corrects for the error under its assumption of classical, non-differential error, and on our data it gives rates between the other two. Accordingly, every per-token rate, share and shift appears under all three treatments, and every other result names its treatment. With every anchor in the scale, the pre-specified treatment gives the slowest decline of the three.

\section{From capability to a price index}
\label{sec:indices}

We compare four indices, and each answers a different question about the same prices. The Jevons index \Eone{} is the unweighted geometric mean of month-on-month price relatives for the endpoints priced in both months. The matched-model chain \Etwo{} follows each endpoint's own price over its life and chains the changes. Both are matched-model indices, and neither prices an entrant until its second month. The pooled time-dummy index \Ethreea{} regresses log prices on capability and month effects over the whole sample. Our headline estimator \Ethreeb{} re-estimates that regression in rolling windows of 13 months and splices the final movement of each window \citep{dehaan2014rollingWindow}. A rolling window lets the price of capability change over time, and in this market it does. At the point estimate with every anchor, the price gradient on capability falls from 1.24 in the first window to 0.63 in the last, in log price per standard deviation of capability. Under plausible values it falls from 0.80 to 0.44. A pooled regression would impose one gradient on all of them.

Within each window the regression is
\begin{equation}
\log p_{et} = \alpha + \beta\,\theta_{e} + \delta_t + \epsilon_{et},
\label{eq:hedonic}
\end{equation}
where $p_{et}$ is the posted price of endpoint $e$ in month $t$. The month effects $\delta_t$ trace the price change at constant capability, and exponentiating them yields the index. A model sold at $k$ reasoning-effort settings shares one posted price across them, so we weight each of its rows by $1/k$ and cluster standard errors on the endpoint. Each spliced movement is the difference between the last two month effects of a window. The two effects are strongly correlated, and we take the variance of their difference from the full cluster-robust covariance before we combine draws (Appendix~\ref{app:estimators}).

The regression also splits every movement into a price term and a quality term. Because it includes month effects, its weighted residuals sum to zero within each month. Differencing the weighted means of two consecutive months gives
\begin{equation}
\delta_t - \delta_{t-1} = \bigl(\overline{\log p}_t - \overline{\log p}_{t-1}\bigr) - \beta\,\bigl(\bar\theta_t - \bar\theta_{t-1}\bigr),
\label{eq:decomp}
\end{equation}
where bars denote weighted means over the endpoints priced in a month. The reported terms satisfy this identity in every window and cell, with a largest residual below $10^{-14}$. Summed over the window, the price term is the change in the mean log posted price of the endpoints on sale, and the quality term is the change in their mean capability valued window by window at the price gradient. The price term does not involve capability, so it takes the same value under every treatment and anchor set.

The headline window runs from January 2025, the base month of the spliced series, to July 2026. We end it in July because August 2026 is the only month that a live pull prices, and the movement of the index in that month measures the change of source. Endpoint keys record their source, and 286 of July's 295 endpoints exit in August while 233 of August's 242 enter. The 203 models priced in both months under any key show a median change of zero in their lowest log price, and 72\% of them keep the same price. We set the end of the window after inspecting the August movement, and in doing so we overruled a rule fixed before that inspection. That rule classed August's entrants by release date and kept the month on the finding that newly observed models carry 12\% of its movement. It tested the wrong failure, since removal shares cannot attribute a movement when the frame turns over this completely. We report its verdict beside our ruling, and we report the window through August 2026 as a sensitivity (Section~\ref{sec:results} and Appendix~\ref{app:window}).

Comparisons across treatments and anchor sets use the 496 models that both capability fits price, 355 of them priced by July 2026. On all 500 models the pre-specified rate is 0.50 (0.11 to 0.87). Intervals come from an endpoint-cluster bootstrap with 4{,}000 resamples, through which the plausible values enter by their draws. We annualise each rate as the cumulative movement of the log index divided by the elapsed months and multiplied by twelve. A rate of 0.51 a year is thus a fall of about 40\% a year, with an interval from 10\% to 59\%, and the matched-model rate of 0.07 is a fall of 7\%. The share of the hedonic decline that matched-model methods miss is one minus the ratio of the \Eone{} rate to the hedonic rate. We print its interval only when the interval of the hedonic rate lies wholly on the faster side of \Eone{}, because the ratio has no stable interval otherwise. For the other cells we give the share of resamples that fall faster than \Eone{}.

The public sources expose no expenditure weights over the whole window, and all our indices are unweighted. Section~\ref{sec:results} reports a weighted check on a subset for 2025. A hedonic index recovers the equilibrium schedule of prices against characteristics \citep{rosen1974hedonic}, and \citet{pakes2003hedonic} shows that it bounds the welfare change from above without measuring it. In a market with subsidised and strategic pricing that schedule may sit far from cost. A deflator is a price-schedule object, and our claims stop there.

\section{Results per token}
\label{sec:results}

\begin{figure}[htb]
  \centering
  \includegraphics[width=\textwidth]{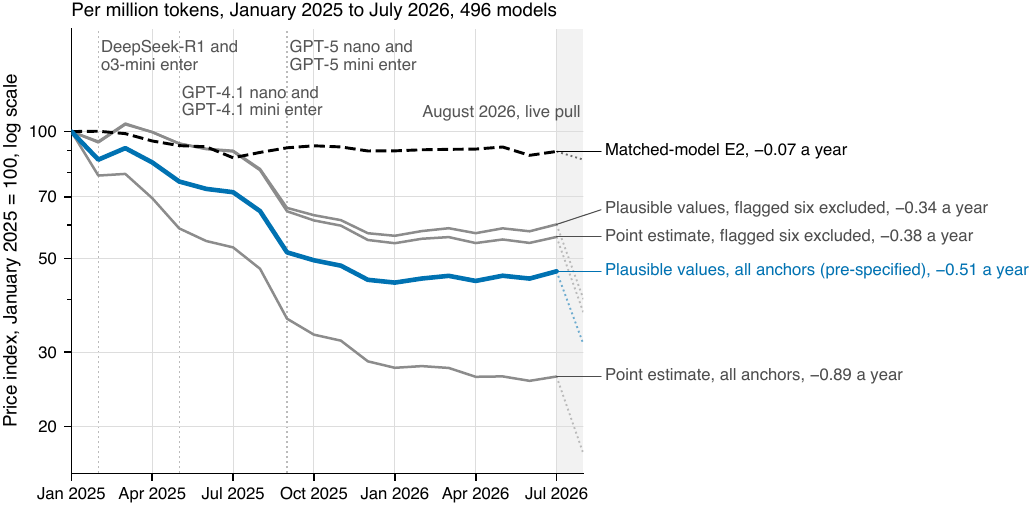}
  \caption{Price indices per million tokens over the headline window for the 496 models that both capability fits price, with January 2025 at 100 on a log scale. The matched-model chain \Etwo{} falls 0.07 a year. The pre-specified hedonic index falls 0.51 a year and the point-estimate index 0.89, while without the six flagged anchors the two fall 0.34 and 0.38. The hedonic indices fall in steps in February, May and September 2025, when the panel first prices the entrants named in the figure. The pre-specified index does not fall from January to July 2026, where its movements sum to $+0.05$ ($-0.19$ to $+0.28$). A model enters in the month the panel first prices it, and for a new release that month can fall one month after its launch. A rule fixed in code before drawing marks a month on three conditions, namely that it is among the three largest movements, that entrants carry more than half of it and that refitting without its new entrants removes more than half of it. A live pull prices August 2026, shaded, and archived captures price every earlier month. The dotted extension into August lies outside the headline window.}
  \label{fig:paths}
\end{figure}

Our main result is that quality-adjusted prices per token fell seven times as fast as matched-model methods record on the pre-specified estimator (Table~\ref{tab:threebytwo} and Figure~\ref{fig:paths}). Over January 2025 to July 2026 the pre-specified index falls 0.51 a year in logs (0.11 to 0.89), while the Jevons index \Eone{} and the matched-model chain \Etwo{} each fall 0.07. Matched-model methods miss 86\% of the decline (37\% to 97\%), and 97.8\% of bootstrap resamples fall faster than \Eone{}. This answers RQ1 in full, with a quality-adjusted annual log fall of 0.51 against a matched-model fall of 0.07. On the endpoints of the 496 models, where \Etwo{} falls 0.08 a year, the gap between the hedonic rate and \Etwo{} is 0.43 (0.01 to 0.82). A deflator that follows matched models would understate the fall in the price of AI inference by that gap each year.

\begin{table}[htb]
\centering
\begin{threeparttable}
\caption{Annual decline in the quality-adjusted price per token, January 2025 to July 2026}
\label{tab:threebytwo}
\small
\setlength{\tabcolsep}{4pt}
\begin{tabular}{@{}lcccccc@{}}
\toprule
 & \multicolumn{3}{c}{All ten anchors} & \multicolumn{3}{c}{Flagged six excluded} \\
\cmidrule(lr){2-4}\cmidrule(l){5-7}
Treatment & Rate & \makecell{Ratio\\to \Etwo} & \makecell{Share\\missed} & Rate & \makecell{Ratio\\to \Etwo} & \makecell{Share\\missed} \\
\midrule
Plausible values & 0.51 & 7.0 & 86\% & 0.34 & 4.7 & 78\% \\
\quad pre-specified & (0.11 to 0.89) & & (37\% to 97\%) & (\mn0.17 to 0.79) & & [84.8\%] \\[3pt]
Regression calibration & 0.75 & 10.4 & 90\% & 0.35 & 4.9 & 79\% \\
 & (0.43 to 1.07) & & (78\% to 97\%) & (\mn0.001 to 0.71) & & [93.8\%] \\[3pt]
Point estimate & 0.89 & 12.3 & 92\% & 0.38 & 5.3 & 81\% \\
 & (0.59 to 1.19) & & (83\% to 98\%) & (0.04 to 0.72) & & [96.0\%] \\
\bottomrule
\end{tabular}
\begin{tablenotes}[flushleft]\footnotesize
\item \textit{Notes}. Rates are annual declines in the log index for the 496 models that both capability fits price, with 95\% intervals from an endpoint-cluster bootstrap of 4{,}000 resamples. A negative bound is a rise. Over the same months \Etwo{} falls 0.07 a year and \Eone{} 0.07. The share missed is one minus the ratio of the \Eone{} rate to the hedonic rate. Its interval appears only when the interval of the rate lies wholly on the faster side of \Eone{}, and square brackets otherwise give the share of resamples that fall faster than \Eone{}. Regression calibration sets the population mean by first priced month. With every anchor, its variants give 0.79 (0.48 to 1.09) for a linear trend and 0.63 (0.31 to 0.96) for one global mean.
\end{tablenotes}
\end{threeparttable}
\end{table}

Rising capability of the models on sale accounts for the whole decline (Table~\ref{tab:decomp}). Over the window the mean log posted price of the endpoints on sale rises 0.07 a year, so the price term is positive. The mean capability of the models on sale rises by 1.46 standard deviations over the same months. Valued window by window at the price gradient, that gain contributes a decline of 0.58 a year under plausible values. Across the cells the gain stays between 1.45 and 1.57 standard deviations, and the gradient that values it carries most of the difference between them. The median gradient across windows is 0.56 in log price per standard deviation under plausible values, 0.82 under regression calibration and 1.00 at the point estimate.

\begin{table}[htb]
\centering
\begin{threeparttable}
\caption{Price and quality terms of the annual rate, January 2025 to July 2026}
\label{tab:decomp}
\small
\setlength{\tabcolsep}{5pt}
\begin{tabular}{@{}llccccc@{}}
\toprule
Anchors & Treatment & \makecell{Price\\term} & \makecell{Quality\\term} & \makecell{Median\\gradient} & \makecell{Capability\\gain} & \makecell{Quality\\term kept} \\
\midrule
All ten & Plausible values & $+0.07$ & $-0.58$ & 0.56 & 1.46 & \\
 & Regression calibration & $+0.07$ & $-0.82$ & 0.82 & 1.45 & \\
 & Point estimate & $+0.07$ & $-0.96$ & 1.00 & 1.48 & \\[3pt]
Flagged six & Plausible values & $+0.07$ & $-0.41$ & 0.42 & 1.57 & 71\% \\
excluded & Regression calibration & $+0.07$ & $-0.43$ & 0.43 & 1.53 & 52\% \\
 & Point estimate & $+0.07$ & $-0.46$ & 0.48 & 1.55 & 47\% \\
\bottomrule
\end{tabular}
\begin{tablenotes}[flushleft]\footnotesize
\item \textit{Notes}. Terms are signed annual changes in logs, and they sum to the negative of the rate in Table~\ref{tab:threebytwo}. The price term is the change in the mean log posted price of the endpoints on sale. The quality term is the change in their mean capability, valued window by window at the price gradient of equation~\eqref{eq:decomp}. The median gradient is in log price per standard deviation of capability across the 18 windows, and the capability gain is the cumulative change in mean capability in standard deviations. The last column gives the quality term without the flagged six as a share of its value with every anchor.
\end{tablenotes}
\end{threeparttable}
\end{table}

The three treatments of measurement error order the rates as Section~\ref{sec:capability} predicts. Plausible values give the slowest of the three rates, and regression calibration falls between plausible values and the point estimate. Regression calibration gives 0.75 (0.43 to 1.07) and the point estimate 0.89 (0.59 to 1.19), against 0.51 for plausible values. The calibration variants give 0.79 (0.48 to 1.09) with a linear trend in the population mean and 0.63 (0.31 to 0.96) with one global mean. With every anchor in the scale all three intervals exclude zero, and matched-model methods miss between 86\% and 92\% of the decline.

The anchor set matters as much as the treatment. Across the three treatments and the two anchor sets the rate runs from 0.34 to 0.89. Four of the six intervals exclude zero, and the two that include it belong to plausible values and regression calibration without the flagged anchors. The gap to \Etwo{} on the same endpoints excludes zero in the three cells with every anchor and includes it in the three without the flagged six. In every cell the central estimate of the hedonic rate exceeds the rate of \Etwo{} by a ratio of 4.7 to 12.3, and matched-model methods miss 78\% to 92\% of the decline. Section~\ref{sec:audit} explains why the anchor set has this much influence.

The decline arrives in steps, in the months when models enter (Figure~\ref{fig:paths} and Table~\ref{tab:timing}). Two monthly movements, in September and February 2025, carry 50\% of the cumulative change in the pre-specified index and 38\% of the change in the point-estimate index with every anchor. In each of the three largest movements the endpoints priced in both months move the log index by at most 0.002 in either direction. Refitting without the month's new releases removes 100\% of the September movement and 86\% of the May movement. February is mixed, with eight of its fifteen entrants newly observed, and refitting without its seven new releases removes 57\% of its movement. Removing one endpoint at a time and summing the shares over each entrant's endpoints attributes 43\% of the September movement to GPT-5 nano and 21\% to GPT-5 mini. The same sums attribute 34\% of the February movement to DeepSeek-R1 and 13\% to o3-mini, and 66\% of the May movement to GPT-4.1 nano and 33\% to GPT-4.1 mini. These sums need not add to the joint removal share. A model enters in the month the panel first prices it, and for a new release that month can fall one month after its launch.

\begin{table}[htb]
\centering
\begin{threeparttable}
\caption{The three largest monthly movements of the pre-specified index}
\label{tab:timing}
\small
\setlength{\tabcolsep}{4pt}
\begin{tabular}{@{}lccccc@{}}
\toprule
Month & Movement & \makecell{Share of\\change} & \makecell{Continuing\\endpoints} & Entrants & \makecell{Removed\\without new} \\
\midrule
September 2025 & $-0.23$ & 29\% & $+0.001$ & 20 and 5 & 100\% \\
February 2025 & $-0.15$ & 20\% & $+0.002$ & 7 and 8 & 57\% \\
May 2025 & $-0.10$ & 14\% & $-0.001$ & 15 and 3 & 86\% \\
\bottomrule
\end{tabular}
\begin{tablenotes}[flushleft]\footnotesize
\item \textit{Notes}. Movements are changes in the log index for the 496 models that both capability fits price. The share of change is each movement's share of the cumulative change over January 2025 to July 2026, and the two largest carry 50\%. Continuing endpoints gives the contribution of endpoints priced in both months. Entrants counts new and newly observed models. A model is new when its release date falls in the entry month or the month before, and newly observed otherwise. The last column gives the share of the movement removed by refitting without the month's new entrants. Removal shares need not add across groups.
\end{tablenotes}
\end{threeparttable}
\end{table}

In its central estimate the pre-specified index completes the whole of its decline in 2025. The eleven movements of February to December 2025 sum to $-0.81$ ($-1.33$ to $-0.22$), and the seven movements of January to July 2026 sum to $+0.05$ ($-0.19$ to $+0.28$). Over those seven months the point-estimate cell with every anchor moves $-0.08$ ($-0.31$ to $+0.14$), and \Etwo{} moves $-0.003$. A matched-model index compares only the prices of continuing endpoints, and those prices barely moved in the months that carry the decline. The gap between the two indices therefore opens in a few entry months, and a matched-model index cannot record those steps by construction.

The choice of window moves the rate, but only within the range that the treatments of measurement error already span. Adding August 2026, the live-pull month, raises the pre-specified rate to 0.73 (0.31 to 1.12) against 0.10 for \Etwo{} over the same months. Starting in February 2025 drops the February step and lowers the rate to 0.43 (0.06 to 0.78), against 0.08 for \Etwo{}. The log-linear slope over the headline window is 0.58 (0.20 to 0.96). Each alternative keeps the hedonic decline faster than \Etwo{}, and each rate depends on which entry months the window contains.

\begin{figure}[htbp]
  \centering
  \includegraphics[width=0.86\textwidth]{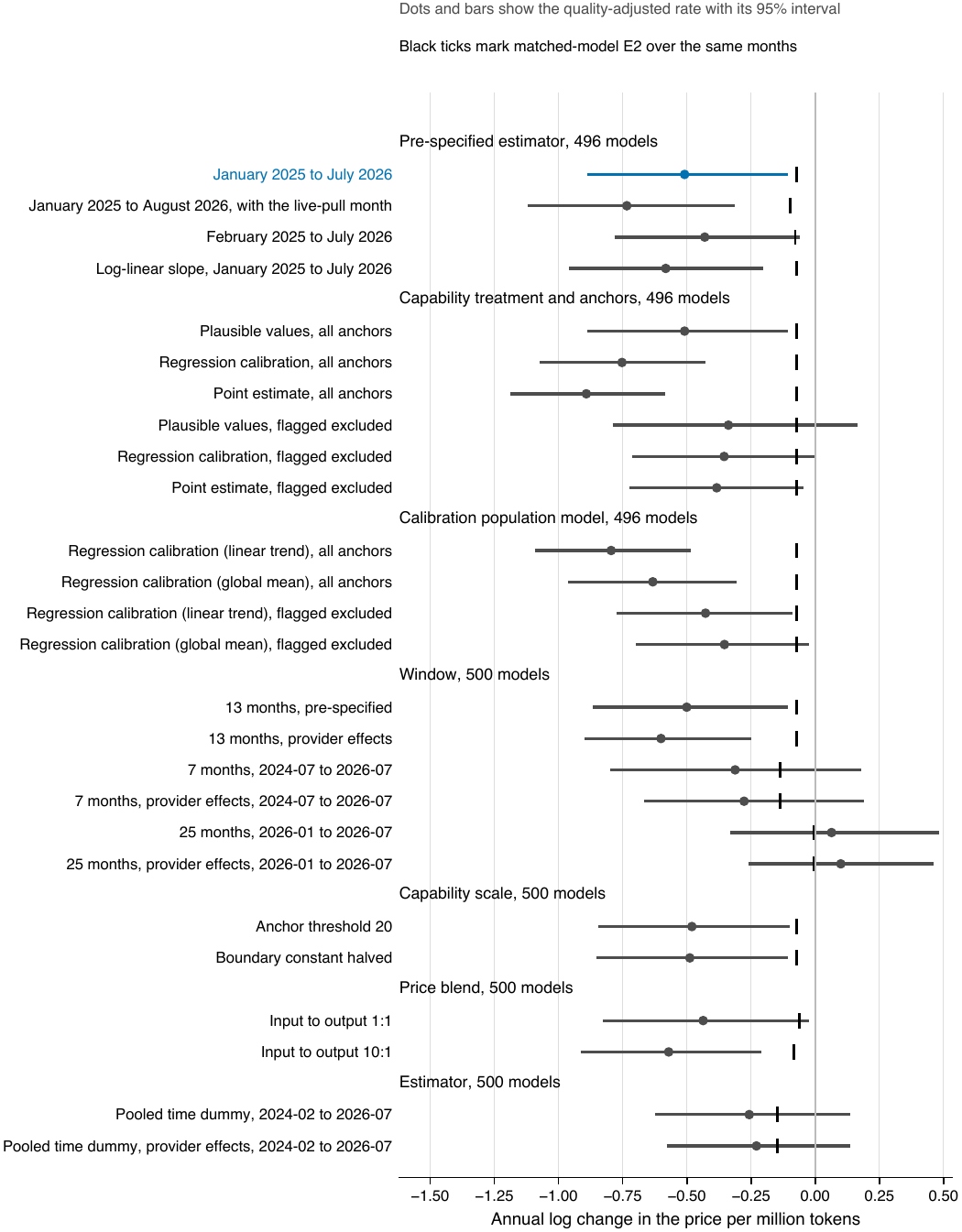}
  \caption{Annual log change in the price per million tokens for every specification, shown with its 95\% interval against matched-model \Etwo{} over the same months (black ticks). Negative values are declines. The first group holds the pre-specified estimator on the headline window and its window sensitivities. The second and third hold the three treatments of measurement error with and without the six flagged anchors and the variants of regression calibration, all on the 496 models that both fits price. The remaining groups hold the pre-registered specifications on all 500 models, with their own windows, scale settings, price blends and estimators. Every specification falls faster than \Etwo{} in its central estimate except the two 25-month windows. Those windows produce movements only from January to July 2026.}
  \label{fig:forest}
\end{figure}

Every pre-registered specification falls faster than \Etwo{} over its own months in its central estimate, except the two 25-month windows (Figure~\ref{fig:forest} and Appendix~\ref{app:specs}). Those windows produce movements only from January to July 2026, a stretch over which neither index establishes a movement. Composition across providers does not drive the decline. Provider fixed effects raise the 13-month rate on all 500 models from 0.50 to 0.60 (0.25 to 0.90). The 7-month windows give 0.31 and 0.28 over July 2024 to July 2026, and the pooled time-dummy index gives 0.26 and 0.23 over February 2024 to July 2026. Their intervals span zero, and over those longer stretches \Etwo{} itself falls 0.14 and 0.15. Price blends of one and ten parts input to one part output give 0.44 and 0.57. A stricter anchor rule and a halved boundary constant give 0.48 and 0.49.

The lower envelope of the panel falls inside the hedonic range. Among models above the January 2025 median of capability, the lowest price per token falls 0.46 a year in logs over the headline window. Above the 75th percentile it falls 0.84 a year. A monthly minimum has no endpoint-bootstrap interval, since resampling endpoints can only remove the cheapest listing or keep it. Across the plausible-value draws the first rate does not move, because the same listings set the monthly minima under every draw, and the second runs from 0.22 to 1.45. At the median over draws the two thresholds admit 14 and 7 models in January 2025 and 268 and 258 in July 2026. The envelope follows one model at a time at the cheap edge of the market, and it answers the narrower question that the threshold studies of Section~\ref{sec:lit} pose.

The OECD's quality-adjusted index fell at a rate inside our range over an earlier period. Its aggregate across modalities, weighted under the OECD's baseline demand scenario, fell about 80\% between March 2023 and January 2025 \citep{andre2025oecdPrices}. That fall amounts to about 0.88 a year in logs, against our range of 0.34 to 0.89. The aggregate covers more modalities than our panel and months that precede our window, so we do not attempt to reconcile the two.

Weighting by usage steepens the hedonic decline in the one subset where we can test it. For 56 models with archived OpenRouter usage in 2025, the pooled hedonic index falls 1.94 a year in logs unweighted (1.05 to 2.80) and 3.07 when weighted by token share (1.70 to 4.06). The matched-model index of the same subset rises under both weightings, by 0.24 as a Jevons index (0.06 to 0.44) and by 0.18 as a T\"ornqvist index ($-0.39$ to 1.00). The subset covers one year, one estimator and the models that one marketplace routes. The check gives the sign of the weighting effect for that subset and leaves its size for the headline unmeasured.

\section{The validity audit}
\label{sec:audit}

The rates of Section~\ref{sec:results} rest on the capability scale, and the scale itself needs an audit. Following \citet{kane2013validity}, we organise that audit as a chain of four inferences from scores to index. Scoring asks whether a score reflects the capability it claims to measure. Generalisation asks whether performance on these benchmarks extends beyond them. Extrapolation asks whether $\theta$ tracks capability outside the tests, and decision asks whether an index built on $\theta$ can bear the use an agency would make of it. Table~\ref{tab:audit} records that scoring fails, generalisation is not established, extrapolation passes and decision passes with its expected sign reversed. The central finding of the audit is that excluding the flagged anchors preserves the ranking of models and changes the spacing on which the index depends.

\begin{table}[htb]
\centering
\begin{threeparttable}
\caption{Dispositions of the pre-registered validity audit}
\label{tab:audit}
\small
\begin{tabularx}{\textwidth}{@{}lX l@{}}
\toprule
Inference & Evidence & Disposition \\
\midrule
Scoring & Six of nine testable anchors show differential functioning by release year after a ceiling control, and seven of ten admit no test by benchmark release & Fail \\
Generalisation & The residual correlation matrix is too sparse to factor, and the weak anchor pairs involve one benchmark & Not established \\
Extrapolation & Capability rises with release date (Spearman 0.880, 757 models) and tracks two external criteria & Pass \\
Decision & Standard errors fall toward the frontier (median 0.187 to 0.084 across quintiles), driven by coverage & Pass, sign reversed \\
\bottomrule
\end{tabularx}
\begin{tablenotes}[flushleft]\footnotesize
\item \textit{Notes}. The pre-registration named each inference and its primary test. Appendix~\ref{app:deviations} records the one instrument we substituted after registration.
\end{tablenotes}
\end{threeparttable}
\end{table}

The scoring inference fails on the evidence of differential functioning. A model trained on a benchmark scores above its capability on that benchmark, and the signature of such exposure is a benchmark that works differently for the models that could have seen it \citep{holland1988mantelHaenszel}. Our registered primary test compares models released before and after each benchmark's publication. Seven of the ten anchors predate nearly every model in the panel and admit no such comparison. The registered secondary test bins models by calendar year of release and controls for ceiling curvature with a quadratic in $\theta$. Six of the nine anchors that it can test show differential functioning. The flags are consistent with contamination, although calendar cohorts differ in more than their exposure to the benchmarks.

\begin{figure}[htb]
  \centering
  \includegraphics[width=\textwidth]{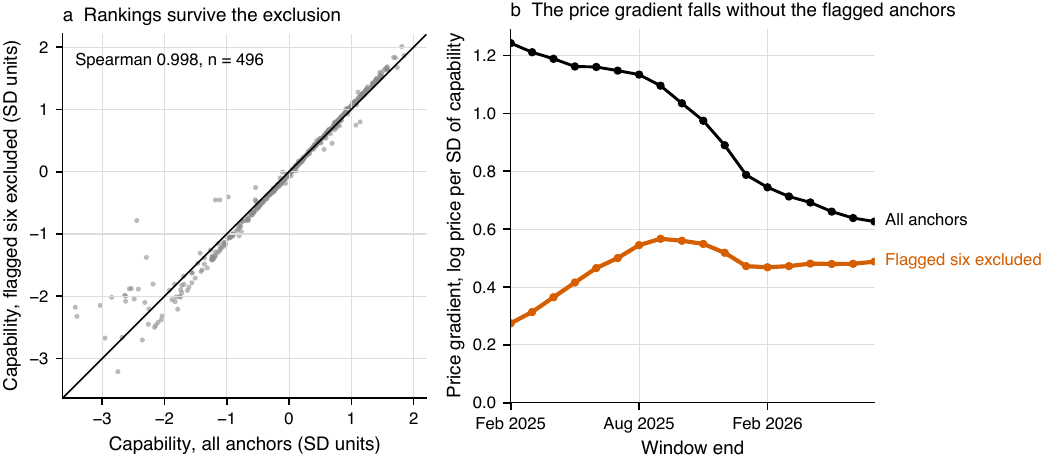}
  \caption{How a ranking-preserving exclusion slows the index. (a) Capability with all ten anchors against capability without the six flagged anchors, for the 496 models that both fits price, in standard deviations. The rankings agree at a Spearman correlation of 0.998. (b) The price gradient on capability, in log price per standard deviation, for each 13-month window by its final month at the point estimate. With every anchor the gradient falls from 1.24 in the first window to 0.63 in the last. Without the flagged six it rises from 0.27 to 0.49 and never exceeds 0.57. The exclusion halves the median gradient and barely moves the capability gain of the models on sale (Table~\ref{tab:decomp}). In that way it slows the index while it leaves the ranking intact.}
  \label{fig:audit}
\end{figure}

Excluding the six flagged anchors leaves the ranking of models intact and slows the index (Figure~\ref{fig:audit}). We refit the scale without them, re-anchor its identification and recompute every index on the 496 models that both fits price. The two capability estimates agree at a Spearman correlation of 0.998. Under regression calibration the exclusion lowers the annual log rate by 0.40 (0.23 to 0.59), and at the point estimate it lowers the rate by 0.51 (0.30 to 0.74). The two calibration variants give shifts of 0.37 (0.21 to 0.55) and 0.28 (0.08 to 0.49). Plausible values damp the shift to 0.17 ($-0.32$ to 0.67), so under the pre-specified treatment we cannot establish it.

The exclusion works through the price gradient on capability (Table~\ref{tab:decomp}). It leaves the price term unchanged by construction, and it barely moves the capability gain of the models on sale, from 1.48 to 1.55 standard deviations at the point estimate. What it does is halve the median gradient, from 1.00 to 0.48 in log price per standard deviation at the point estimate and from 0.82 to 0.43 under regression calibration. The quality term without the flagged six accordingly falls to 47\% and 52\% of its value with every anchor. A hedonic index consumes the spacing of the capability scale through this gradient, and agreement on ranks places no bound on the spacing. Agreement on rankings therefore does not license claims about levels or increments.

Plausible values damp the shift because they attenuate the gradient most where it is steepest. Their median gradient is 0.56 with every anchor and 0.42 without the flagged six, against 1.00 and 0.48 at the point estimate. Under plausible values the quality term accordingly keeps 71\% of its value without the flagged anchors. One caveat applies to the whole exercise. The exclusion list comes from our own audit, so the sensitivity cannot corroborate the flags that produced it. What it measures is how much the index depends on six benchmarks that the scale uses to link its two eras and whose integrity the registered test could not settle.

The generalisation inference is not established, and the data leave open whether capability is one dimension or several. Different benchmarks score different subsets of models, and the matrix of residual correlations is too sparse to factor. Rankings agree across anchor pairs at a median of 0.809, yet the weak pairs all involve GSM8K against the older knowledge benchmarks, a pattern consistent with a partly distinct construct. Saturation is easier to measure. A model-free instrument records the share of frontier models that score within $\varepsilon$ of a benchmark's ceiling, and by that instrument three benchmarks saturate within the panel at $\varepsilon = 0.05$. The count runs from one at $\varepsilon = 0.02$ to six at $\varepsilon = 0.10$.

The extrapolation inference passes, and it is the least informative outcome in the chain. Capability rises with release date at a Spearman correlation of 0.880 across 757 dated models. It tracks LMArena's rank of human preference at a Spearman correlation of $-0.700$ across 69 models, where the best model holds rank one \citep{lmarena2026}. It tracks Epoch's capability index at 0.930 across 504 models \citep{epoch2026eci}. Almost any competent capability measure would pass these checks. A stricter check at the level of individual questions fails. On five benchmarks with readable question-level logs, a two-parameter item response model fitted within each benchmark agrees with $\theta$ at a median rank correlation of 0.728. The agreement reaches 0.7 on three of the five and falls to 0.098 on chess puzzles.

The decision inference passes with its expected sign reversed, and the reversal moves the warning about reliability to the other end of the market. We expected saturation to make capability least precise at the frontier. In fact the posterior standard error falls toward the frontier, from a median of 0.187 in the lowest quintile of capability to 0.084 in the highest. By release year it falls from 0.197 for models released in 2024 to 0.097 for those released in 2026. Coverage by benchmarks drives the pattern. Conditioning on the number of scored cells per model raises the explained variance of the standard error from 0.126 to 0.690. The standard error also tracks the inverse square root of each model's test information at a Spearman correlation of 0.975. Labs and leaderboards evaluate frontier models on more benchmarks, and our scale therefore measures the sparsely evaluated back catalogue worst.

The answer to RQ2 is partial, and it turns on the anchor set. With every anchor in the scale the decline survives all three treatments of measurement error, and its size depends on the treatment. Without the flagged anchors the rate falls to between 0.34 and 0.38, still 4.7 to 5.3 times the rate of \Etwo{}, and only the interval at the point estimate excludes zero.

\section{The price per completed task}
\label{sec:pertask}

The data cannot answer RQ3. None of the twelve per-task intervals of rates and wedges excludes zero, and this section explains why the per-task series are so uninformative. The per-task price multiplies a model's posted price by the output tokens it uses on a fixed evaluation suite, and we run the hedonic regression of Section~\ref{sec:indices} on that price. Because a spliced window produces movements only once it has filled, the three per-task series start in different months. Table~\ref{tab:pertask} reports each series under plausible values and at the point estimate, together with the per-task \Etwo{} over the same months and the wedge between \Etwo{} and the hedonic rate.

\begin{table}[htb]
\centering
\begin{threeparttable}
\caption{Annual change in the quality-adjusted price per completed task}
\label{tab:pertask}
\small
\setlength{\tabcolsep}{4pt}
\begin{tabular}{@{}llccc@{}}
\toprule
Series and months & Treatment & Hedonic rate & \makecell{Per-task\\\Etwo} & Wedge \\
\midrule
Pooled \Ethreea & Plausible values & $+0.31$ ($-0.61$ to $+1.45$) & $+0.06$ & $-0.25$ ($-1.33$ to $+0.56$) \\
\quad 2025-04 to 2026-07 & Point estimate & $+0.27$ ($-0.62$ to $+1.36$) & $+0.06$ & $-0.21$ ($-1.24$ to $+0.58$) \\[3pt]
\Ethreeb, 7 months & Plausible values & $+0.79$ ($-0.16$ to $+1.75$) & $+0.09$ & $-0.71$ ($-1.56$ to $+0.16$) \\
\quad 2025-09 to 2026-07 & Point estimate & $+0.75$ ($-0.18$ to $+1.66$) & $+0.09$ & $-0.66$ ($-1.49$ to $+0.19$) \\[3pt]
\Ethreeb, 13 months & Plausible values & $-0.48$ ($-1.45$ to $+0.62$) & $-0.23$ & $+0.25$ ($-0.85$ to $+1.20$) \\
\quad 2026-03 to 2026-07 & Point estimate & $-0.58$ ($-1.52$ to $+0.46$) & $-0.23$ & $+0.35$ ($-0.71$ to $+1.25$) \\
\bottomrule
\end{tabular}
\begin{tablenotes}[flushleft]\footnotesize
\item \textit{Notes}. Rates are signed annual changes in logs with 95\% endpoint-cluster bootstrap intervals, and a negative rate is a fall. The wedge is the per-task \Etwo{} rate minus the hedonic rate. Each token count dates from July or August 2026 and applies to every earlier month. The 13-month window is the pre-specified estimator, and its base month is March 2026.
\end{tablenotes}
\end{threeparttable}
\end{table}

\begin{figure}[htb]
  \centering
  \includegraphics[width=0.78\textwidth]{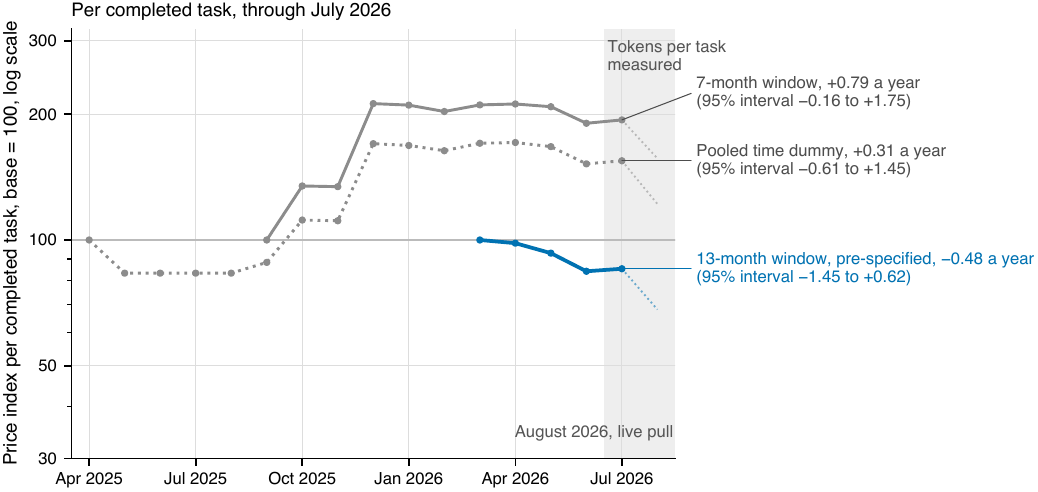}
  \caption{Price indices per completed task through July 2026, each on its own base of 100 and on a log scale. The pooled time-dummy index rises 0.31 a year from April 2025, and the 7-month window rises 0.79 from September 2025. The pre-specified 13-month window falls 0.48 from March 2026. Every interval spans zero (Table~\ref{tab:pertask}). Artificial Analysis measured tokens per task in July or August 2026, the shaded months. We carry each count back to every earlier month. The dotted extensions into August 2026 cross the change of source and lie outside the headline window.}
  \label{fig:pertask}
\end{figure}

Every interval in Table~\ref{tab:pertask} spans zero, and the three hedonic rates that Figure~\ref{fig:pertask} draws do not even agree in sign. The pooled index rises 0.31 a year from April 2025, with an interval that runs from a fall of 0.61 to a rise of 1.45. The 7-month window rises 0.79 from September 2025, between a fall of 0.16 and a rise of 1.75. The pre-specified 13-month window falls 0.48 from March 2026, between a fall of 1.45 and a rise of 0.62. The wedges against the per-task \Etwo{} span zero as well. On these data the per-task series cannot distinguish a rise from a fall.

The token counts are the main reason. Artificial Analysis measured them once, in July or August 2026, and we carry each count back over the 16 per-task months through July 2026. As a result, 66\% of the per-task observations carry dates before July 2026 and rest on a measurement taken later. A series built this way holds each model's token use fixed at its value in the summer of 2026, and it cannot see any change in a model's own token use over its life.

The same bound limits comparison with studies of the price of thought. \citet{emberson2026plungingPrice} estimate that the cost of reaching a given level of performance has fallen about thirteen times a year since 2023, taking the cheapest route to each level across models, reasoning settings and token budgets. Our per-task prices cover 14 of the 87 models on their cost frontier over April 2025 to July 2026, and 54 of the 578 model-months on that frontier. Our per-task series therefore cannot test their decline.

\section{Limitations}
\label{sec:limits}

Seven limitations qualify these conclusions, and we state them in decreasing order of consequence. First, choices of measurement move the rate, and the data cannot rank those choices. The treatment of measurement error and the anchor set together move the rate between 0.34 and 0.89. We do not know which treatment is correct on these data, and our own audit supplies the exclusion list that defines one of the two anchor sets. These choices leave intact the finding that the hedonic index falls faster than \Etwo{} in the central estimate of every cell. They leave the size of the decline uncertain beyond the width of any one interval.

Second, the headline window is short and its decline is lumpy. Two of its 18 monthly movements carry half of the cumulative change. In its last seven months the index does not fall, and its movements over those months sum to $+0.05$ ($-0.19$ to $+0.28$). Starting the window one month later lowers the rate to 0.43. A rate fitted to such a window describes these months and supports no forecast of later ones.

Third, we observe posted list prices only, and transaction prices may have moved differently. If discounts below list grew over the window, transaction prices fell faster than our index. If discounts shrank, transaction prices fell more slowly. The public record does not tell us which. Fourth, our indices carry no expenditure weights. The one weighted check covers 56 models in 2025 under a pooled estimator, and in that check token-share weights steepen the subset's hedonic decline from 1.94 to 3.07 a year. The check gives the sign of the weighting effect for that subset and leaves its size for the headline unmeasured.

Fifth, every per-task series holds each model's token use at its measurement in July or August 2026. No such series can show change in token use over a model's life, and every per-task interval spans zero. Sixth, the registered primary test of contamination cannot reach seven of the ten anchors, because they predate nearly every model in the panel. The early years of the equated scale rest on those ten anchors, for which model developers reported 40\% of the scored cells, and no later public record can replace those scores. The audit shows how much the index depends on the flagged anchors. It cannot show that the anchors which pass are clean.

Seventh, the panel covers one market in one period, namely text and code inference sold through APIs. Image, audio and video generation lie outside it, and so do consumer subscriptions and models used inside other products.

Finally, we did not pre-register several of the analyses reported here. We added them in revision after the first public version. They are regression calibration with the comparison across treatments, the decomposition into price and quality terms, the timing of the decline, the lower envelope, the OECD comparison and the usage check. We also moved the end of the headline window after inspecting the August 2026 movement. Appendix~\ref{app:deviations} records these changes beside the earlier deviations from the pre-registration. Every number we report recomputes with one command from the released provenance ledgers and archived page captures, on the same code with fixed seeds. That command does not re-pull the original web pages, so a reader reproduces the hand checks of Section~\ref{sec:data} by inspection.

\section{Conclusion}
\label{sec:conclusion}

We find that the quality-adjusted price of AI inference per token fell from January 2025 to July 2026 several times faster than matched-model methods record. Quality enters this market through new models priced close to the models they surpass. A matched-model index follows continuing models and misses the steps that entry produces. A hedonic index on a capability scale records them. Our pre-specified index falls 0.51 a year in logs against 0.07 for matched-model methods, and the whole of that decline comes from the capability term.

The size of the decline depends on the capability scale beneath the index. A hedonic index inherits the spacing of its scale, and our audit shows that this spacing depends on the anchors. Excluding six flagged anchors leaves the ranking of models intact. It lowers the annual log rate by 0.40 to 0.51 under regression calibration and at the point estimate, and under the pre-specified treatment the slowdown is smaller and its interval includes zero. The reason is that a price index consumes the distances between models on the scale, and rankings place no bound on those distances. Agreement on rankings therefore does not license claims about levels or increments.

Three practices follow for measuring prices in this market. The quality characteristic should be a latent scale whose measurement error enters every published statistic, and the published index should carry its range across treatments of that error. The audit of the scale should test its spacing as well as its rankings. A per-task price needs token counts measured month by month before it can say whether the price of completed work fell. Benchmark-based adjustment from public data is feasible now. Its result rests on choices that a statistical agency would need to publish.

\ifblind
\section*{Data and code availability}

We release the dataset, its codebook, the crosswalk, every analysis file with the hashes of its inputs, the provenance ledgers behind a committed checksum manifest and the code that recomputes every reported number. Version 1.0.2 of the release is archived under a DOI given on the title page, with data under CC BY 4.0 and code under the MIT licence.
\else
\section*{Data and code availability}

{\emergencystretch=1em\relax %
We release the dataset, its codebook, the crosswalk, every analysis file with the hashes of its inputs, the provenance ledgers behind a committed checksum manifest and the code that recomputes every reported number. We archive version 1.0.2 of the release under DOI \datasetdoi{}, with data under CC BY 4.0 and code under the MIT licence. The same record links the pre-registration and the public log of every deviation from it.\par}

\section*{Funding and competing interests}

The author received no external funding for this research and declares no competing interests.
\fi

\section*{Use of AI assistance}

\aidisclosure{}

\section*{Note on versions}

Version 1 of this paper reported a quality-adjusted decline of 0.73 a year in logs over January 2025 to August 2026, with 87\% of it invisible to matched-model methods. August 2026 is the only month priced from a live pull of aggregator pages. Every earlier month comes from archived captures, and the index's fall in that month measures the change of source. The 203 models priced in both July and August show a median price change of zero. This version ends the headline window in July 2026, on which the pre-specified estimator falls 0.51 a year in logs (95\% interval 0.11 to 0.89) against 0.07 for matched-model methods. It reports the full window as a sensitivity and the index under three treatments of measurement error in capability. It states the basis of the contamination sensitivity, and it withdraws the per-task result of version 1 as one that the data here do not establish.

\bibliographystyle{apalike}
\bibliography{refs}

\clearpage
\appendix
\counterwithin*{table}{section}
\renewcommand{\thetable}{\thesection\arabic{table}}
\counterwithin*{figure}{section}
\renewcommand{\thefigure}{\thesection\arabic{figure}}

\section{Glossary}
\label{app:glossary}

The glossary in Table~\ref{tab:glossary} defines the terms, abbreviations and notation of the main text.

\begin{table}[htb]
\centering
\begin{threeparttable}
\caption{Terms, abbreviations and notation}
\label{tab:glossary}
\small
\begin{tabularx}{\textwidth}{@{}lX@{}}
\toprule
Term & Meaning \\
\midrule
API & Application programming interface, the channel through which providers sell inference \\
Endpoint & One model's listing on one provider's channel, and the unit on which we cluster \\
Headline window & January 2025 to July 2026, 18 monthly movements of the spliced index \\
Live pull & Prices read from current pages in August 2026, the only month not priced from archived captures \\
\Eone & Jevons index, the unweighted geometric mean of matched month-on-month price relatives \\
\Etwo & Matched-model chain, following each endpoint's own price over its life \\
\Ethreea & Pooled time-dummy hedonic index over the whole sample \\
\Ethreeb & Rolling-window time-dummy hedonic index with 13-month windows, spliced, and the headline estimator \\
$\theta_m$ & Capability of model $m$ on the equated scale, in standard deviations where stated \\
$a_k$, $d_k$ & Discrimination and difficulty of benchmark $k$ in equation~\eqref{eq:crm} \\
$\beta$ & Price gradient on capability, in log price per standard deviation of capability \\
$\delta_t$ & Month effect of the hedonic regression in equation~\eqref{eq:hedonic} \\
Anchor & One of ten benchmarks fitted first and held fixed to link benchmark generations \\
Flagged six & The anchors that show differential functioning under the registered secondary test \\
Plausible values & Twenty draws of $\theta$ per model from the estimation posterior, the pre-specified treatment \\
Point estimate & The estimate $\hat\theta_m$ used as if it were known \\
Regression calibration & The estimate replaced by its expected true value given the estimate, equation~\eqref{eq:rc} \\
Price term & Change in the mean log posted price of the endpoints on sale, equation~\eqref{eq:decomp} \\
Quality term & Change in the mean capability of the endpoints on sale, valued at the price gradient \\
Share missed & One minus the ratio of the \Eone{} rate to the hedonic rate \\
Wedge & Per-task \Etwo{} rate minus the per-task hedonic rate \\
Log rate & Annual change in the natural logarithm of an index, so that a fall of $x$ a year is a fall of $1 - e^{-x}$ a year \\
\bottomrule
\end{tabularx}
\end{threeparttable}
\end{table}

\section{Estimators added in revision}
\label{app:estimators}

The decomposition of Section~\ref{sec:indices} follows from the normal equations of the window regression. Let each endpoint's rows carry weights that sum to one within a month. Because equation~\eqref{eq:hedonic} includes a month effect, its weighted residuals sum to zero within each month. Hence the weighted mean log price of month $t$ equals $\alpha + \beta\bar\theta_t + \delta_t$, and differencing two months of one window gives equation~\eqref{eq:decomp}. Summed over the headline window, the price terms telescope to the change in the mean log price of the endpoints on sale between January 2025 and July 2026. The quality terms do not telescope, because each movement carries its own window's gradient. The weight is $1/k$ per endpoint-month, and each endpoint-month therefore sums to one. The largest absolute residual of the identity over all windows and cells lies below $10^{-14}$.

Regression calibration needs two inputs beyond the capability fit, namely the population mean $\mu_m$ and the variance $\tau^2$ of true capability around it. Each model's standard error $s_m$ comes from the fit, and each calibrated value enters equation~\eqref{eq:hedonic} in place of $\hat\theta_m$. The primary variant takes $\mu_m$ as the mean of $\hat\theta$ among the models first priced in the same month. The two others take it from an ordinary least squares line in the ordinal of that month and from one global mean. We estimate $\tau^2$ by the method of moments as the variance of $\hat\theta_m - \mu_m$ less the mean of $s_m^2$ over the index models, floored at zero. It takes 0.275, 0.376 and 0.825 with every anchor. The calibration runs once before the resampling loop, and the bootstrap holds the calibrated values fixed.

Each spliced movement is $m_t = \hat\delta_t - \hat\delta_{t-1}$ within one window. Its variance is $\operatorname{Var}(\hat\delta_t) + \operatorname{Var}(\hat\delta_{t-1}) - 2\operatorname{Cov}(\hat\delta_t, \hat\delta_{t-1})$, taken from the cluster-robust covariance with clusters on endpoints. Dropping the covariance overstates the standard error of a movement several times over and can produce bounds that imply negative index levels. The bootstrap resamples endpoints with replacement 4{,}000 times under seed 20261103, and the plausible values come from seed 8102026. Each resample recomputes every rate, ratio, share and shift. In this way every contrast carries the covariance of its two parts.

\section{The headline window}
\label{app:window}

The ruling on the window turned on what the August 2026 movement measures. The live pull prices August 2026 alone, and it supplies 385 of the month's 430 rows in the index frame. Endpoint keys record their source, and 286 of July's 295 endpoints exit in August while 233 of August's 242 enter, with 9 priced in both months. July had priced 210 of the 357 model identifiers behind August's entrant rows under another key. The 203 models priced in both months under any key show a median change of zero in their lowest log price, and 72\% of them keep the same price. Removal shares cannot attribute a movement when the frame turns over this completely, and the share of the movement attributed to all entrants together is $-102\%$. The rule we fixed before the inspection classed entrants by release date, found that newly observed models carry 12\% of the movement and kept the month. We report that verdict, and Table~\ref{tab:window} reports the window sensitivities on both samples.

\begin{table}[htb]
\centering
\begin{threeparttable}
\caption{Window sensitivities of the pre-specified estimator}
\label{tab:window}
\footnotesize
\setlength{\tabcolsep}{4pt}
\begin{tabular}{@{}lcccc@{}}
\toprule
 & \multicolumn{2}{c}{496 models} & \multicolumn{2}{c}{500 models} \\
\cmidrule(lr){2-3}\cmidrule(l){4-5}
Window & Rate & \Etwo{} & Rate & \Etwo{} \\
\midrule
January 2025 to July 2026, headline & 0.51 (0.11 to 0.89) & 0.07 & 0.50 (0.11 to 0.87) & 0.07 \\
January 2025 to August 2026, with the live pull & 0.73 (0.31 to 1.12) & 0.10 & 0.73 (0.32 to 1.11) & 0.10 \\
February 2025 to July 2026 & 0.43 (0.06 to 0.78) & 0.08 & 0.43 (0.07 to 0.77) & 0.08 \\
Log-linear slope, headline months & 0.58 (0.20 to 0.96) & 0.07 & 0.55 (0.18 to 0.92) & 0.07 \\
\bottomrule
\end{tabular}
\begin{tablenotes}[flushleft]\footnotesize
\item \textit{Notes}. Annual declines in the log index with 95\% bootstrap intervals, under plausible values with every anchor. \Etwo{} is the matched-model chain over the same months.
\end{tablenotes}
\end{threeparttable}
\end{table}

\section{Specifications}
\label{app:specs}

Fifteen pre-registered specifications lie behind Figure~\ref{fig:forest}, and Table~\ref{tab:specs} sets each against \Etwo{} over its own months at its own price blend. The first twelve use all 500 models, and the last three re-estimate cells of Table~\ref{tab:threebytwo} on the 496 models that both fits price. Figure~\ref{fig:forest} shows those three only once, as cells. Thirteen of the fifteen fall faster than \Etwo{} in their central estimates, and the two exceptions are the 25-month windows.

\begin{table}[htb]
\centering
\begin{threeparttable}
\caption{Pre-registered specifications}
\label{tab:specs}
\footnotesize
\setlength{\tabcolsep}{4pt}
\begin{tabular}{@{}llcc@{}}
\toprule
Specification & Months & Rate & \Etwo{} \\
\midrule
\Ethreeb, 13 months, headline & 2025-01 to 2026-07 & 0.50 (0.11 to 0.87) & 0.07 \\
\Ethreeb, 13 months, provider effects & 2025-01 to 2026-07 & 0.60 (0.25 to 0.90) & 0.07 \\
\Ethreeb, 7 months & 2024-07 to 2026-07 & 0.31 ($-0.18$ to 0.80) & 0.14 \\
\Ethreeb, 7 months, provider effects & 2024-07 to 2026-07 & 0.28 ($-0.19$ to 0.67) & 0.14 \\
\Ethreeb, 25 months & 2026-01 to 2026-07 & $-0.06$ ($-0.48$ to 0.33) & 0.01 \\
\Ethreeb, 25 months, provider effects & 2026-01 to 2026-07 & $-0.10$ ($-0.46$ to 0.26) & 0.01 \\
Anchor threshold 20 & 2025-01 to 2026-07 & 0.48 (0.10 to 0.84) & 0.07 \\
Boundary constant halved & 2025-01 to 2026-07 & 0.49 (0.10 to 0.85) & 0.07 \\
Price blend 1:1 & 2025-01 to 2026-07 & 0.44 (0.03 to 0.83) & 0.06 \\
Price blend 10:1 & 2025-01 to 2026-07 & 0.57 (0.21 to 0.91) & 0.08 \\
\Ethreea, pooled & 2024-02 to 2026-07 & 0.26 ($-0.14$ to 0.62) & 0.15 \\
\Ethreea, pooled, provider effects & 2024-02 to 2026-07 & 0.23 ($-0.14$ to 0.58) & 0.15 \\[3pt]
Flagged six excluded, point estimate & 2025-01 to 2026-07 & 0.38 (0.04 to 0.72) & 0.07 \\
All anchors, point estimate & 2025-01 to 2026-07 & 0.89 (0.59 to 1.19) & 0.07 \\
Flagged six excluded, plausible values & 2025-01 to 2026-07 & 0.34 ($-0.17$ to 0.79) & 0.07 \\
\bottomrule
\end{tabular}
\begin{tablenotes}[flushleft]\footnotesize
\item \textit{Notes}. Annual declines in the log index with 95\% bootstrap intervals, where a negative rate is a rise. The rows use plausible values with every anchor unless they state otherwise. \Etwo{} is the matched-model chain over the row's months at the row's price blend. Four rows have counterparts in the joint bootstrap of Table~\ref{tab:threebytwo}, and two of them differ from it at the second decimal. The two bootstraps share a seed and a universe of endpoints and consume the random stream differently. Their percentiles therefore differ at the second decimal, at 0.12 to 0.86 against 0.11 to 0.87 for the headline and 0.59 to 1.17 against 0.59 to 1.19 for the point estimate. We print the joint bootstrap's intervals, since every shift, share and gap in the main table comes from its resamples.
\end{tablenotes}
\end{threeparttable}
\end{table}

\section{Deviations from the pre-registration}
\label{app:deviations}

Three elements changed after registration and before the affected estimation ran. Our public log records each change at the time it occurred. First, we rescaled to proportions the four benchmarks that report percentages and excluded two benchmarks that we could not classify. Together the two changes left 4{,}605 usable cells of the 4{,}672 registered. Second, the homoscedastic specification makes statistical information constant in capability, and information curves cannot date saturation under it. We replaced it with the model-free measure of ceiling proximity and kept genuine information curves for the five benchmarks with item-level data. Third, the registered nominal series named its publisher's annual report, and the figure we used came from the same publisher's mid-year update of the same market model.

Three further changes came in revision, after estimation. First, the headline window ends in July 2026 (Section~\ref{sec:indices} and Appendix~\ref{app:window}). We made that ruling after inspecting the August 2026 movement. The window of version 1 ran to August 2026 and appears here as a sensitivity. Second, the comparisons across treatments use the 496 models that both capability fits price, and the registered sample of 500 models gives 0.50 (0.11 to 0.87). Third, this version omits the registered illustration that deflated a survey estimate of enterprise spending on model APIs \citep{menlo2025llmMarket}. That survey differs in scope from our panel, and the gap between the two rates already states the yearly error in a deflator that follows matched models. Section~\ref{sec:limits} lists the analyses that we added in revision without registration.

\section{Material carried from version 1}
\label{app:carried}

We carry over below the appendix of version 1, on the frozen panel of February 2024 to August 2026. We omit its tables that reported rates on that window, since Tables~\ref{tab:threebytwo} to~\ref{tab:pertask} and Appendices~\ref{app:window} and~\ref{app:specs} carry their July-window equivalents. Its sections cover the estimator, the archival recovery and verification of the panel, the benchmark battery, the anchor rule, the audit chain, the index detail and the reproduction procedure.

\providecommand{\appendixtablesize}{\scriptsize}
\providecommand{\appendixtabcolsep}{6pt}
\providecommand{\Description}[1]{}
\captionsetup{hypcap=false}

\subsection{Mathematical basis}
\label{app:math}

Supports Section~\ref{sec:capability}.

We derive here each estimator the paper uses and record why the chosen form beats its nearest alternatives on these data. Documented degeneracies forced several of the choices, and the forcing is itself evidence.

\paragraph{The continuous response model and its two failed alternatives.} Scores are proportions, and the natural first likelihood is a beta density with mean linked to $a_k(\theta_m - d_k)$. On this table that likelihood has no maximum. Scores on 86 per cent of benchmarks fall on coarse grids and multiple models tie at identical proportions, and the beta precision parameter then diverges with the likelihood unbounded. The second natural form keeps a Gaussian residual on the logit scale and gives each benchmark its own variance $\sigma_k^2$. This form fails on thin benchmarks. A benchmark scored on few models can drive $\hat{\sigma}_k \to 0$ around an interpolating fit and again unbound the likelihood, and the thinnest benchmark in the table carries five scored cells. We therefore place one shared residual scale on logit scores,
\begin{equation}
\operatorname{logit}(s_{mk}) = a_k(\theta_m - d_k) + \epsilon_{mk}, \qquad \epsilon_{mk} \sim \mathcal{N}(0, \sigma^2),
\end{equation}
the homoscedastic form of Samejima's continuous response model. We estimate it by maximum likelihood over $(\theta, a, d, \sigma)$ jointly, with GPQA Diamond pinned at $a = 1$, $d = 0$ for identification. Before the logit we move the 89 cells scored at exactly 0 or 1 inward by 0.001. The shared $\sigma$ carries a cost the audit exploits. Under this form the statistical information about $\theta$ is constant in $\theta$, and saturation claims require the separate instrument of section~\ref{app:audit}.

\begin{table}[htbp]
  \centering
  \small
  \resizebox{\ifdim\width>\linewidth\linewidth\else\width\fi}{!}{\begin{tabular}{p{0.19\linewidth}p{0.26\linewidth}p{0.3\linewidth}l}
\toprule
Specification & Degeneracy & Conditioning observed & Outcome \\
\midrule
Beta likelihood on proportions & Precision unbounded above; a thinly scored benchmark is fitted almost exactly & $e^{24.8}$ precision reached; negative Hessian eigenvalue at condition number $5\times10^{42}$ & no maximum \\
Per-benchmark residual SD & $\sigma_k$ collapses toward zero on a thin benchmark & objective unbounded below & no maximum \\
Standardised $\theta$\newline in objective & every model's ability depends on every other's; gradient explodes & not conditioned & no stable optimum \\
\textbf{Ratified: logit scale, shared $\sigma$, reference item} & one $\sigma$ cannot collapse; $a_k\theta_m$ pinned at one anchor & min eigenvalue 0.0861, condition number 2.18e+05, 0 negative & \textbf{converged} \\
\bottomrule
\end{tabular}
}
  \caption{Evidence for the two likelihood degeneracies that forced the homoscedastic form. For the beta likelihood the profile increases without bound as precision diverges on grid-valued benchmarks. For per-benchmark variances the thinnest benchmarks drive their residual scale toward zero around an interpolating fit. The shared-scale model of section~\ref{app:math} is thus the least structured form the data permit, a forced choice reported as such.}
  \label{tab:b1-degeneracy}
\end{table}

\paragraph{Two-stage fixed-parameter equating.} Benchmarks retire and new ones appear. No single benchmark spans the panel, and benchmarks that overlap two generations must carry the scale across them. Stage one fits the model on the ten anchor benchmarks alone over the 480 models they reach, and this stage fixes the anchors' $(a_k, d_k)$ on one scale. Stage two holds those parameters fixed and estimates every remaining benchmark's parameters and every model's $\theta$ against them. Joint estimation in one pass would let thin recent benchmarks bend the scale that links the eras. Fixing the anchors makes the linkage explicit and auditable, and the DIF check of section~\ref{app:audit} can attack it. The price is that anchor defects propagate by construction. For that reason the audit runs on the anchors first, and the contamination sensitivity re-anchors the scale, since re-weighting alone would leave the defects in place.

\paragraph{Plausible values and Rubin combination.} A downstream statistic computed at $\hat{\theta}$ alone treats an estimate as truth and understates its own uncertainty. We draw $L = 20$ values $\theta^{(\ell)}$ per model from the estimation posterior, compute every downstream quantity $Q^{(\ell)}$ once per draw and combine with Rubin's rules. The point estimate is $\bar{Q} = L^{-1}\sum_\ell Q^{(\ell)}$ and its variance is $V = \bar{W} + (1 + L^{-1})\,B$, where $\bar{W}$ is the mean within-draw sampling variance and $B$ the between-draw variance of $Q^{(\ell)}$. We report the decomposition wherever it is informative. For the version-1 headline movements the between-draw share is small, a median 4.9 per cent of movement variance, and price sampling dominates the intervals.

\paragraph{Movement variance by contrast, and the naive error it prevents.} A spliced movement is $m_t = \hat{\delta}_t - \hat{\delta}_{t-1}$ inside one window regression. Its variance is
\begin{equation}
\operatorname{Var}(m_t) = \operatorname{Var}(\hat{\delta}_t) + \operatorname{Var}(\hat{\delta}_{t-1}) - 2\operatorname{Cov}(\hat{\delta}_t, \hat{\delta}_{t-1}).
\end{equation}
We take the covariance from the cluster-robust covariance matrix, clustered on priced endpoints. Month coefficients within a window share the same capability slope estimate and much of the same sample, and their covariance is large and positive as a result. An implementation that reads only coefficient standard errors drops it. Dropping it inflates the movement standard error by a factor of 7.6 at the median here, and in our first implementation it produced interval bounds that implied negative index levels. Levels cumulate movements across overlapping windows whose covariances the splice does not track. We report level standard errors as indicative and quote movement intervals in the text.

\paragraph{Splicing and the elapsed-month basis.} Each 13-month window contributes the movement between its two most recent months. The spliced series chains these movements from the first window's end. The series' base month is thus the last month of the first window, and a $k$-month window consumes $k - 1$ months before its first movement. The per-token base falls at 2025-01 and the per-task base at 2026-03. Annualised rates divide cumulated log movements by elapsed calendar months between the base and the endpoint, never by the count of index points. Every table states its base, because two spliced series with different windows do not share one.

\paragraph{Endpoint-cluster bootstrap for cross-unit contrasts.} The divergence and the wedge are differences between rates estimated on overlapping samples. Their variances need the covariance between the two rates, and the per-unit regressions do not supply it. The bootstrap resamples priced endpoints with replacement, since the endpoint is the independent sampling unit under the clustering already in use. It recomputes both spliced rates on each resample, forms the contrast within the resample and repeats $B = 4000$ times. Capability uncertainty enters through the plausible values. Resampling endpoints jointly for both units carries the cross-unit covariance. The interval accordingly narrows relative to treating the rates as independent exactly when their correlation is positive. We drop and count resamples in which a unit's spliced series cannot be formed, 2.2 per cent of draws for the divergence. The dropped resamples are the thinnest ones, and the reported intervals are hence mildly optimistic. We state that direction wherever the margin is small.

\paragraph{Differential functioning with a ceiling control.} For anchor $k$ and cohort indicator $c_m$, the test compares the fitted model by likelihood ratio against an augmented one in which cohort shifts the benchmark's location, $\operatorname{logit}(s_{mk}) = a_k(\theta_m - d_k - \gamma_k c_m) + \epsilon_{mk}$. Saturated benchmarks generate apparent $\gamma_k$ from curvature alone. Near the ceiling equal capability gaps produce compressed score gaps, and cohorts differ in mean capability. The control augments the null with a quadratic in $\theta$ for the benchmark before the cohort term enters. We report only flags that survive the control. The primary cohort split, before and after the benchmark's own release, matches the contamination mechanism and is testable on three of the ten anchors. The secondary split by calendar year is testable on nine and supplies the operative evidence.

\paragraph{The ceiling-proximity saturation measure.} For benchmark $k$ and month $t$, let $F_t$ be the top decile of $\theta$ among models released by $t$. The measure is the share of $F_t$ whose score on $k$ lies within $\varepsilon = 0.05$ of the ceiling of 1, and a benchmark counts as saturated in the first month the share reaches one half. The measure is model-free, and it does not inherit the homoscedastic form's silence about saturation. We report its threshold sensitivity at $\varepsilon \in \{0.02, 0.10\}$. The saturated count moves from one to six across that range, and the status of MATH Level 5 does not move.

\subsection{Panel construction and provenance}
\label{app:panel}

Supports Section~\ref{sec:data}.

\begin{figure}[htbp]
  \centering
  \includegraphics[width=\linewidth]{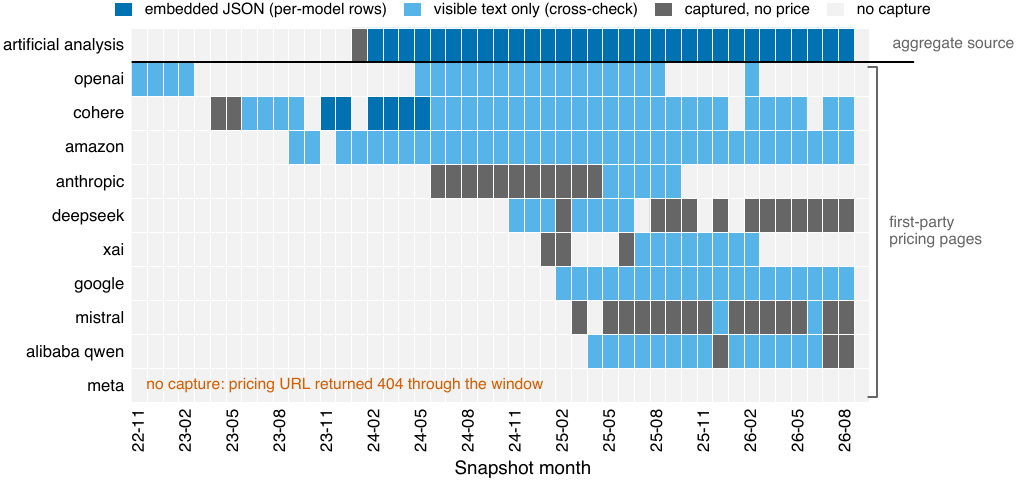}
  \caption{Archival price coverage by provider and month. The aggregate source row sits above the rule because it carries every provider's prices, and the rows below it form the first-party cross-check tier. Dark cells mark months recovered from embedded application state, light cells months where only rendered text survives, grey cells captures without readable prices, and the annotated empty row a pricing URL the archive never captured. Coverage thins toward the market's beginning. The panel start accordingly follows the pre-registered three-provider rule, and the earlier months enter only the descriptive prologue.}
  \Description{Archival price coverage by provider and month. The aggregate source row sits above the rule because it carries every provider's prices, and the rows below it form the first-party cross-check tier. Dark cells mark months recovered from embedded application state, light cells months where only rendered text survives, grey cells captures without readable prices, and the annotated empty row a pricing URL the archive never captured. Coverage thins toward the market's beginning. The panel start accordingly follows the pre-registered three-provider rule, and the earlier months enter only the descriptive prologue.}
  \label{fig:fa1}
\end{figure}

The panel's claim to accuracy rests on procedure, and this section records it. Archival recovery proceeded in two passes, a text pass over rendered pages and a structural pass over the JSON application state embedded in archived pages. The second pass recovered 13{,}829 price observations that no visible text carries, including the aggregate source's full per-model history. Figure~\ref{fig:fa1} maps the result by provider and month. The meta row is empty because the provider's pricing URL returns a redirect the archive never captured, a gap recorded with its reason like every other.

\begin{table}[htbp]
  \centering
  \small
  \resizebox{\ifdim\width>\linewidth\linewidth\else\width\fi}{!}{\begin{tabular}{lrr}
\toprule
Recovery outcome & Cells & Share \\
\midrule
Provider-months in the recovery frame & 517 & -- \\
Observed, either carrier & 177 & 34.2\% \\
\quad of which embedded JSON state (per-model rows) & 37 & 20.9\% \\
\quad of which visible text only (cross-check) & 140 & 79.1\% \\
Captured but no price in either carrier & 47 & 9.1\% \\
No Wayback capture in the month & 281 & 54.4\% \\
Not reached & 12 & 2.3\% \\
Providers in the frame & 11 & -- \\
\bottomrule
\end{tabular}
}
  \caption{Archival recovery statistics by pass. The structural pass over embedded application state recovers the aggregate source's full per-model history, and no rendered table carries that history. The text pass survives where only rendered pages exist. The two passes together determine the coverage grid of Figure~\ref{fig:fa1}.}
  \label{tab:d1-wayback}
\end{table}

\paragraph{Verification and adjudication.} We hand-checked forty price observations against their recorded source URLs, and all forty matched without correction. Table~\ref{tab:d2-verification} reproduces them. We drew the hand-checked rows from the most recent month and from the two current-page sources, and they do not cover the archival recovery pass. We release a 5\% random sample of all observed rows, 1{,}051 rows spanning every month, for inspection alongside them. Separately, we adjudicated row by row the twenty largest divergences between the aggregate source and the cross-check tier. Twelve resolved as third-party resellers hosting a model at their own price. Five resolved as stale listings of retired endpoints, converted to recorded exits under a ruled exit rule. The remainder resolved as channel differences between a provider's own venues. We record each disposition beside its row.

\begin{table}[htbp]
  \centering
  \small
  \resizebox{\ifdim\width>\linewidth\linewidth\else\width\fi}{!}{\begin{tabular}{lr}
\toprule
Verification sample & Rows \\
\midrule
Rows drawn for hand-checking & 40 \\
Verified against the source page & 40 \\
Corrections required & 0 \\
\quad drawn from artificial\_analysis & 24 \\
\quad drawn from openrouter & 16 \\
\bottomrule
\end{tabular}
}
  \caption{Counts by category for the forty hand-checked rows. We checked each against its recorded source URL, and all forty matched without correction. The rows come from the most recent month and from the two current-page sources and do not cover the archival recovery pass. A 5\% random sample of all observed rows, 1{,}051 rows spanning every month, ships for inspection alongside them. The row-level sample ships with the released dataset with each observation carrying its source URL and archive timestamp, and the accuracy claim is inspectable there in full.}
  \label{tab:d2-verification}
\end{table}

\begin{table}[htbp]
  \centering
  \small
  \resizebox{\ifdim\width>\linewidth\linewidth\else\width\fi}{!}{\begin{tabular}{llp{0.34\linewidth}r}
\toprule
Ruled class & Models & Review mark as recorded & Median gap \\
\midrule
retired endpoint & 5 & flag\_first\_party (5) & 35.0\% \\
channel price discrimination & 2 & flag\_first\_party (2) & 60.3\% \\
expected third-party spread & 13 & expected\_third\_party (12), needs\_manual\_provider\_check (1) & 48.4\% \\
\textbf{All flagged} & \textbf{20} & -- & 48.4\% \\
\bottomrule
\end{tabular}
}
  \caption{Adjudication of the twenty largest divergences between the aggregate source and the cross-check tier. Most resolve as third-party resellers hosting a model at their own price. The remainder resolve as stale listings of retired endpoints, converted to recorded exits under the ruled exit rule, or as channel differences within one provider. Each row carries its disposition. We resolve no divergence silently.}
  \label{tab:d3-divergences}
\end{table}

\begin{table}[htbp]
  \centering
  \small
  \resizebox{\ifdim\width>\linewidth\linewidth\else\width\fi}{!}{\begin{tabular}{lr}
\toprule
Gap-row count, as defined & Rows \\
\midrule
In the ratified window 2024-02 to 2026-08, 31 months & 601 \\
Adding 2026-09, one month beyond the ratified end & 612 \\
All months in the frozen panel, prologue included & 763 \\
\midrule \multicolumn{2}{l}{\itshape Reasons, in-window} \\ \\
\quad endpoint\_retired & 226 \\
\quad no posted price on the Artificial Analysis model record & 190 \\
\quad no Wayback capture in this month & 122 \\
\quad archived page carries embedded JSON state but no input/output pric... & 27 \\
\quad archived page carries no price in the HTML or in embedded JSON & 17 \\
\quad not fetched in this run (cache-only) & 12 \\
\quad other reasons (3 kinds) & 7 \\
\bottomrule
\end{tabular}
}
  \caption{Gap accounting over the panel window, with each definition stated beside its count. We record gaps with reasons and never interpolate them. Retired endpoints count as exits. We define here the in-window count the main text cites, against the neighbouring definitions a reader might otherwise conflate.}
  \label{tab:d4-gaps}
\end{table}

\subsection{Benchmark battery}
\label{app:battery}

Supports Section~\ref{sec:capability}.

\noindent\begin{minipage}{\linewidth}
\captionof{table}{The sixty-four benchmarks of the frozen score table. Role records whether a benchmark anchors the equating. Cells counts scored models. Provenance classifies who produced each score, with developer-reported mass concentrating in the older anchors. Saturation dates come from the ceiling-proximity measure of section~\ref{app:math} at $\varepsilon = 0.05$, blank where a benchmark never saturates in-window. Each benchmark's release date carries one source URL in the repository's metadata file, and the released dataset records the canonical reference or documentation page for every benchmark in the battery.}
\label{tab:e1-battery}
\end{minipage}
{\appendixtablesize\setlength{\tabcolsep}{\appendixtabcolsep}\begin{longtable}{lrrlllll}
\toprule
Benchmark & Cells & Models & Provenance & Released & Scale & Role & Saturates \\
\midrule
\endfirsthead
\toprule
Benchmark & Cells & Models & Provenance & Released & Scale & Role & Saturates \\
\midrule
\endhead
\midrule
\multicolumn{8}{r}{\itshape continued} \\
\endfoot
\bottomrule
\endlastfoot
adversarial\_nli\_external & 15 & 11 & lab\_reported & 2019-10-31 & proportion &  & -- \\
aider\_polyglot\_external & 72 & 71 & third\_party\_leaderboard & 2024-12-21 & percentage &  & -- \\
apex\_agents\_external & 55 & 55 & unattributed & -- & proportion &  & -- \\
arc\_agi\_2\_external & 172 & 158 & unattributed & 2025-05-17 & proportion &  & -- \\
arc\_agi\_external & 188 & 168 & third\_party\_leaderboard & 2019-11-05 & proportion &  & -- \\
arc\_ai2\_external & 134 & 77 & lab\_reported & 2018-03-14 & proportion & anchor & -- \\
balrog\_external & 36 & 35 & third\_party\_leaderboard & 2024-11-20 & proportion &  & -- \\
bbh\_external & 88 & 50 & lab\_reported & 2022-10-17 & proportion &  & -- \\
blueprint\_bench\_2\_external & 21 & 21 & third\_party\_leaderboard & -- & proportion &  & -- \\
bool\_q\_external & 136 & 77 & lab\_reported & 2019-05-24 & proportion &  & -- \\
btf3\_external & 9 & 7 & third\_party\_leaderboard & -- & proportion &  & -- \\
cad\_eval\_external & 15 & 15 & third\_party\_leaderboard & 2025-04-22 & proportion &  & -- \\
chess\_puzzles & 153 & 153 & epoch & -- & proportion &  & -- \\
cl\_bench\_external & 23 & 23 & unattributed & -- & proportion &  & -- \\
cl\_bench\_life\_external & 17 & 17 & unattributed & -- & proportion &  & -- \\
common\_sense\_qa\_2\_external & 6 & 6 & third\_party\_leaderboard & 2022-01-14 & proportion &  & -- \\
critpt\_external & 139 & 139 & unattributed & -- & proportion &  & -- \\
cursorbench\_external & 31 & 31 & third\_party\_leaderboard & -- & proportion &  & -- \\
cybench\_external & 22 & 22 & third\_party\_leaderboard & 2024-08-15 & proportion &  & -- \\
deepresearchbench\_external & 36 & 36 & third\_party\_leaderboard & 2025-06-13 & proportion &  & -- \\
deepswe\_external & 50 & 50 & third\_party\_leaderboard & -- & proportion &  & -- \\
enigma\_eval\_external & 46 & 46 & third\_party\_leaderboard & 2025-02-13 & proportion &  & -- \\
forecastbench\_external & 75 & 75 & unattributed & 2024-09-30 & percentage &  & -- \\
frontiercode\_external & 20 & 20 & third\_party\_leaderboard & -- & proportion &  & -- \\
frontiermath & 101 & 101 & epoch & 2024-11-07 & proportion &  & -- \\
frontiermath\_tier\_4 & 72 & 72 & epoch & -- & proportion &  & -- \\
gbaeval\_external & 23 & 23 & third\_party\_leaderboard & -- & proportion &  & -- \\
gdp\_pdf\_external & 21 & 21 & third\_party\_leaderboard & -- & proportion &  & -- \\
gdpval\_external & 11 & 11 & unattributed & 2025-09-25 & proportion &  & -- \\
gpqa\_diamond & 255 & 255 & epoch & 2023-11-20 & proportion & anchor & -- \\
gsm8k\_external & 170 & 93 & lab\_reported & 2021-10-27 & proportion & anchor & -- \\
gso\_external & 38 & 38 & third\_party\_leaderboard & 2025-05-29 & proportion &  & -- \\
hella\_swag\_external & 113 & 76 & lab\_reported & 2019-05-19 & proportion & anchor & 2023-05 \\
hle\_external & 51 & 46 & unattributed & 2025-01-24 & proportion &  & -- \\
lambada\_external & 53 & 26 & lab\_reported & 2016-06-20 & proportion &  & -- \\
live\_bench\_external & 54 & 52 & third\_party\_leaderboard & 2024-06-27 & percentage &  & -- \\
math\_level\_5 & 108 & 108 & epoch & 2021-03-05 & proportion & anchor & 2025-05 \\
metr\_time\_horizons\_external & 50 & 50 & third\_party\_leaderboard & 2025-03-18 & proportion & anchor & -- \\
mindcube\_external & 5 & 5 & third\_party\_leaderboard & -- & proportion &  & -- \\
mmlu\_external & 217 & 136 & lab\_reported & 2020-09-07 & proportion & anchor & -- \\
mystery\_game\_puzzles & 47 & 47 & epoch & -- & proportion &  & -- \\
open\_book\_qa\_external & 70 & 42 & lab\_reported & 2018-09-08 & proportion &  & -- \\
os\_world\_external & 20 & 9 & third\_party\_leaderboard & 2024-04-11 & percentage &  & -- \\
osworld\_2\_external & 10 & 8 & third\_party\_leaderboard & -- & proportion &  & -- \\
otis\_mock\_aime\_2024\_2025 & 230 & 230 & epoch & 2024-12-19 & proportion & anchor & 2026-04 \\
piqa\_external & 113 & 60 & lab\_reported & 2019-11-26 & proportion &  & -- \\
posttrainbench\_external & 35 & 32 & unattributed & -- & proportion &  & -- \\
proofbench\_external & 52 & 52 & third\_party\_leaderboard & -- & proportion &  & -- \\
rli\_external & 12 & 12 & unattributed & -- & proportion &  & -- \\
scicode\_external & 129 & 128 & third\_party\_leaderboard & 2024-07-18 & proportion &  & -- \\
science\_qa\_external & 26 & 23 & third\_party\_leaderboard & 2022-09-20 & proportion &  & -- \\
simplebench\_external & 91 & 91 & third\_party\_leaderboard & 2024-10-31 & proportion &  & -- \\
simpleqa\_verified & 71 & 71 & epoch & -- & proportion &  & -- \\
spatialviz\_bench\_external & 8 & 8 & third\_party\_leaderboard & 2025-07-10 & proportion &  & -- \\
superglue\_external & 10 & 8 & lab\_reported & 2019-05-02 & proportion &  & -- \\
surface\_evolver\_bench\_external & 24 & 24 & third\_party\_leaderboard & -- & proportion &  & -- \\
swe\_bench\_verified & 35 & 33 & epoch & 2024-08-13 & proportion &  & -- \\
terminalbench\_external & 202 & 59 & third\_party\_leaderboard & 2025-05-19 & proportion &  & -- \\
the\_agent\_company\_external & 16 & 14 & third\_party\_leaderboard & 2024-12-18 & proportion &  & -- \\
trivia\_qa\_external & 115 & 39 & lab\_reported & 2017-05-09 & proportion &  & -- \\
video\_mme\_external & 50 & 50 & third\_party\_leaderboard & 2024-05-31 & proportion &  & -- \\
vpct\_external & 38 & 38 & third\_party\_leaderboard & 2025-01-30 & proportion &  & -- \\
weirdml\_external & 163 & 162 & third\_party\_leaderboard & 2025-01-16 & proportion & anchor & -- \\
wino\_grande\_external & 137 & 80 & lab\_reported & 2019-07-24 & proportion & anchor & -- \\
\end{longtable}
}
\addtocounter{table}{-1}

\subsection{Anchor selection}
\label{app:anchors}

Supports Section~\ref{sec:capability}.

\begin{figure}[htbp]
  \centering
  \includegraphics[width=\linewidth]{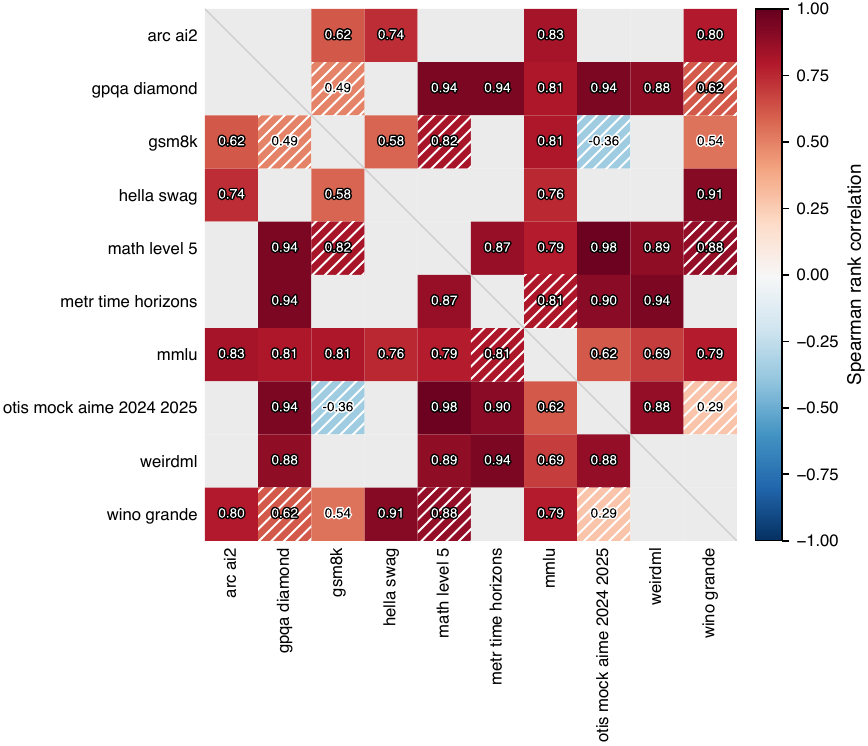}
  \caption{Cross-anchor rank consistency over shared models. Each cell is the Spearman correlation of model rankings between two anchor benchmarks. Hatching marks pairs sharing fewer than twenty models, and such pairs are evidence in neither direction. The single negative pair joins GSM8K to a recent mathematics benchmark on eleven shared models. The weak reliable pairs all involve GSM8K against the older knowledge anchors. The generalisation inference reads that pattern as suggestive of more than one dimension without settling it.}
  \Description{Cross-anchor rank consistency over shared models. Each cell is the Spearman correlation of model rankings between two anchor benchmarks. Hatching marks pairs sharing fewer than twenty models, and such pairs are evidence in neither direction. The single negative pair joins GSM8K to a recent mathematics benchmark on eleven shared models. The weak reliable pairs all involve GSM8K against the older knowledge anchors. The generalisation inference reads that pattern as suggestive of more than one dimension without settling it.}
  \label{fig:fa2}
\end{figure}

We reproduce here the anchor rule and everything it turned on. A benchmark qualifies with at least ten scored models on each side of 1 July 2024 and a post-boundary score standard deviation of at least 0.10. Table~\ref{tab:f1-anchors} reports both quantities for every candidate, and a reader can move the thresholds. Five candidates sit within one model of the count threshold, two binding on the early side and three on the late. Figure~\ref{fig:fa2} reports rank consistency between pairs of anchors over their shared models.

\begin{table}[htbp]
  \centering
  \small
  \resizebox{\ifdim\width>\linewidth\linewidth\else\width\fi}{!}{\begin{tabular}{lrrrrrrrll}
\toprule
Anchor & Cells pre & Cells post & Models pre & Models post & SD post & At ceiling & Shared & Provenance & Clears 20 \\
\midrule
gpqa\_diamond & 38 & 217 & 38 & 217 & 0.163 & 0.114 & 1778 & epoch & yes \\
otis\_mock\_aime\_2024\_2025 & 21 & 209 & 21 & 209 & 0.296 & 0.222 & 1715 & epoch & yes \\
weirdml\_external & 10 & 151 & 10 & 150 & 0.188 & 0.018 & 1136 & third\_party\_leaderboard & no \\
mmlu\_external & 177 & 38 & 102 & 33 & 0.117 & 0.000 & 786 & lab\_reported & yes \\
math\_level\_5 & 34 & 74 & 34 & 74 & 0.229 & 0.204 & 763 & epoch & yes \\
wino\_grande\_external & 126 & 11 & 69 & 11 & 0.110 & 0.000 & 506 & lab\_reported & no \\
arc\_ai2\_external & 123 & 10 & 66 & 10 & 0.219 & 0.045 & 500 & lab\_reported & no \\
metr\_time\_horizons\_external & 11 & 38 & 11 & 38 & 0.119 & 0.000 & 478 & third\_party\_leaderboard & no \\
hella\_swag\_external & 97 & 10 & 60 & 10 & 0.129 & 0.018 & 461 & lab\_reported & no \\
gsm8k\_external & 146 & 16 & 69 & 16 & 0.180 & 0.035 & 437 & lab\_reported & no \\
\bottomrule
\end{tabular}
}
  \caption{Every anchor candidate against the selection rule, with cell and distinct-model counts on each side of the boundary and the post-boundary score standard deviation. Five candidates sit within one model of the count threshold, two binding on the early side and three on the late. The table supplies the means to move the thresholds and see what the list becomes.}
  \label{tab:f1-anchors}
\end{table}

\subsection{Audit evidence}
\label{app:audit}

Supports Section~\ref{sec:audit}.

\phantomsection\label{app:audit-detail}

We collect here the audit's quantitative evidence. It comprises the differential-functioning results for all ten anchors under both cohort designs, the item-level information curves for the five benchmarks with readable logs and the agreement between those item-level fits and the capability estimate. Section~\ref{sec:audit} states what each establishes.

\noindent\begin{minipage}{\linewidth}
\captionof{table}{Differential-functioning results for the ten anchors under both cohort designs, with and without the ceiling-curvature control. The primary pre-and-post-release design is untestable for seven of ten anchors for want of any pre-release comparison group, the finding the main text elevates. The calendar-year design flags six of the nine testable anchors after the control. The flags form the contamination-exclusion list whose index consequences Section~\ref{sec:results} quantifies.}
\label{tab:g1-dif}
\end{minipage}
{\appendixtablesize\setlength{\tabcolsep}{\appendixtabcolsep}\begin{longtable}{llrrrrrl}
\toprule
Benchmark & Status & $n$ & Groups & $p$ uniform & $p$ total & $p$ curv.-ctrl &  \\
\midrule
\endfirsthead
\toprule
Benchmark & Status & $n$ & Groups & $p$ uniform & $p$ total & $p$ curv.-ctrl &  \\
\midrule
\endhead
\midrule
\multicolumn{8}{r}{\itshape continued} \\
\endfoot
\bottomrule
\endlastfoot
\multicolumn{8}{l}{\itshape Anchors, pre/post benchmark release (registered primary)} \\
\quad weirdml\_external & tested & 161 & 25/136 & 0.5708 & 0.0009 & 0.0000 & flagged \\
\quad metr\_time\_horizons\_external & tested & 49 & 21/28 & 0.0047 & 0.0161 & 0.0263 & flagged \\
\quad otis\_mock\_aime\_2024\_2025 & tested & 230 & 48/182 & 0.0997 & 0.2475 & 0.0000 & flagged \\
\quad arc\_ai2\_external & too\_thin & 133 & 0/133 & -- & -- & -- &  \\
\quad gpqa\_diamond & too\_thin & 255 & 7/248 & -- & -- & -- &  \\
\quad gsm8k\_external & too\_thin & 162 & 1/161 & -- & -- & -- &  \\
\quad hella\_swag\_external & too\_thin & 107 & 0/107 & -- & -- & -- &  \\
\quad math\_level\_5 & too\_thin & 108 & 0/108 & -- & -- & -- &  \\
\quad mmlu\_external & too\_thin & 215 & 0/215 & -- & -- & -- &  \\
\quad wino\_grande\_external & too\_thin & 137 & 0/137 & -- & -- & -- &  \\
\multicolumn{8}{l}{\itshape Anchors, calendar year (secondary)} \\
\quad mmlu\_external & tested & 215 & 132/83 & 0.0000 & 0.0000 & 0.0005 & flagged \\
\quad wino\_grande\_external & tested & 137 & 102/35 & 0.0000 & 0.0000 & 0.0000 & flagged \\
\quad arc\_ai2\_external & tested & 133 & 106/27 & 0.0194 & 0.0000 & 0.0008 & flagged \\
\quad gpqa\_diamond & tested & 255 & 161/94 & 0.0039 & 0.0001 & 0.0002 & flagged \\
\quad hella\_swag\_external & tested & 107 & 84/23 & 0.0001 & 0.0003 & 0.0004 & flagged \\
\quad gsm8k\_external & tested & 162 & 132/30 & 0.0064 & 0.0008 & 0.0011 & flagged \\
\quad math\_level\_5 & tested & 108 & 63/45 & 0.0234 & 0.0731 & 0.0588 &  \\
\quad otis\_mock\_aime\_2024\_2025 & tested & 230 & 137/93 & 0.5017 & 0.1962 & 0.5117 &  \\
\quad weirdml\_external & tested & 161 & 95/66 & 0.7525 & 0.8344 & 0.8893 &  \\
\quad metr\_time\_horizons\_external & too\_thin & 49 & 43/6 & -- & -- & -- &  \\
\multicolumn{8}{l}{\itshape Anchors, developer, unstratified} \\
\quad otis\_mock\_aime\_2024\_2025 & tested & 230 & 158/72 & 0.0031 & 0.0019 & 0.0019 & flagged \\
\quad metr\_time\_horizons\_external & tested & 50 & 26/24 & 0.2891 & 0.0582 & 0.1778 &  \\
\quad hella\_swag\_external & tested & 94 & 74/20 & 0.2663 & 0.0665 & 0.0398 & flagged \\
\quad weirdml\_external & tested & 160 & 109/51 & 0.0303 & 0.0938 & 0.0885 &  \\
\quad gpqa\_diamond & tested & 255 & 180/75 & 0.0526 & 0.1019 & 0.0479 & flagged \\
\quad mmlu\_external & tested & 192 & 145/47 & 0.2988 & 0.1159 & 0.2600 &  \\
\quad math\_level\_5 & tested & 108 & 79/29 & 0.1562 & 0.3424 & 0.3952 &  \\
\quad gsm8k\_external & tested & 139 & 93/46 & 0.2389 & 0.4915 & 0.3614 &  \\
\quad wino\_grande\_external & tested & 120 & 94/26 & 0.3406 & 0.6058 & 0.5960 &  \\
\quad arc\_ai2\_external & tested & 108 & 75/33 & 0.5150 & 0.7900 & 0.9410 &  \\
\multicolumn{8}{l}{\itshape Anchors, developer, within Epoch-run stratum} \\
\quad arc\_ai2\_external & no\_cells & -- & -- & -- & -- & -- &  \\
\quad gsm8k\_external & no\_cells & -- & -- & -- & -- & -- &  \\
\quad hella\_swag\_external & no\_cells & -- & -- & -- & -- & -- &  \\
\quad metr\_time\_horizons\_external & no\_cells & -- & -- & -- & -- & -- &  \\
\quad mmlu\_external & no\_cells & -- & -- & -- & -- & -- &  \\
\quad weirdml\_external & no\_cells & -- & -- & -- & -- & -- &  \\
\quad wino\_grande\_external & no\_cells & -- & -- & -- & -- & -- &  \\
\quad otis\_mock\_aime\_2024\_2025 & tested & 230 & 158/72 & 0.0031 & 0.0019 & 0.0019 & flagged \\
\quad gpqa\_diamond & tested & 255 & 180/75 & 0.0526 & 0.1019 & 0.0479 & flagged \\
\quad math\_level\_5 & tested & 108 & 79/29 & 0.1562 & 0.3424 & 0.3952 &  \\
\multicolumn{8}{l}{\itshape All other benchmarks, pre/post benchmark release} \\
\quad apex\_agents\_external & no\_dates & -- & -- & -- & -- & -- &  \\
\quad blueprint\_bench\_2\_external & no\_dates & -- & -- & -- & -- & -- &  \\
\quad btf3\_external & no\_dates & -- & -- & -- & -- & -- &  \\
\quad chess\_puzzles & no\_dates & -- & -- & -- & -- & -- &  \\
\quad cl\_bench\_external & no\_dates & -- & -- & -- & -- & -- &  \\
\quad cl\_bench\_life\_external & no\_dates & -- & -- & -- & -- & -- &  \\
\quad critpt\_external & no\_dates & -- & -- & -- & -- & -- &  \\
\quad cursorbench\_external & no\_dates & -- & -- & -- & -- & -- &  \\
\quad deepswe\_external & no\_dates & -- & -- & -- & -- & -- &  \\
\quad frontiercode\_external & no\_dates & -- & -- & -- & -- & -- &  \\
\quad frontiermath\_tier\_4 & no\_dates & -- & -- & -- & -- & -- &  \\
\quad gbaeval\_external & no\_dates & -- & -- & -- & -- & -- &  \\
\quad gdp\_pdf\_external & no\_dates & -- & -- & -- & -- & -- &  \\
\quad mindcube\_external & no\_dates & -- & -- & -- & -- & -- &  \\
\quad mystery\_game\_puzzles & no\_dates & -- & -- & -- & -- & -- &  \\
\quad osworld\_2\_external & no\_dates & -- & -- & -- & -- & -- &  \\
\quad posttrainbench\_external & no\_dates & -- & -- & -- & -- & -- &  \\
\quad proofbench\_external & no\_dates & -- & -- & -- & -- & -- &  \\
\quad rli\_external & no\_dates & -- & -- & -- & -- & -- &  \\
\quad simpleqa\_verified & no\_dates & -- & -- & -- & -- & -- &  \\
\quad surface\_evolver\_bench\_external & no\_dates & -- & -- & -- & -- & -- &  \\
\quad arc\_agi\_2\_external & tested & 172 & 27/145 & 0.3838 & 0.0000 & 0.0003 & flagged \\
\quad forecastbench\_external & tested & 75 & 20/55 & 0.0221 & 0.0000 & 0.8866 &  \\
\quad balrog\_external & tested & 36 & 16/20 & 0.1325 & 0.3218 & 0.1670 &  \\
\quad adversarial\_nli\_external & too\_thin & 15 & 0/15 & -- & -- & -- &  \\
\quad aider\_polyglot\_external & too\_thin & 72 & 14/58 & -- & -- & -- &  \\
\quad arc\_agi\_external & too\_thin & 188 & 0/188 & -- & -- & -- &  \\
\quad bbh\_external & too\_thin & 88 & 0/88 & -- & -- & -- &  \\
\quad bool\_q\_external & too\_thin & 123 & 0/123 & -- & -- & -- &  \\
\quad cad\_eval\_external & too\_thin & 14 & 14/0 & -- & -- & -- &  \\
\quad common\_sense\_qa\_2\_external & too\_thin & 3 & 0/3 & -- & -- & -- &  \\
\quad cybench\_external & too\_thin & 22 & 6/16 & -- & -- & -- &  \\
\quad deepresearchbench\_external & too\_thin & 35 & 11/24 & -- & -- & -- &  \\
\quad enigma\_eval\_external & too\_thin & 46 & 9/37 & -- & -- & -- &  \\
\quad frontiermath & too\_thin & 101 & 7/94 & -- & -- & -- &  \\
\quad gdpval\_external & too\_thin & 11 & 7/4 & -- & -- & -- &  \\
\quad gso\_external & too\_thin & 38 & 12/26 & -- & -- & -- &  \\
\quad hle\_external & too\_thin & 51 & 7/44 & -- & -- & -- &  \\
\quad lambada\_external & too\_thin & 53 & 0/53 & -- & -- & -- &  \\
\quad live\_bench\_external & too\_thin & 52 & 6/46 & -- & -- & -- &  \\
\quad open\_book\_qa\_external & too\_thin & 70 & 0/70 & -- & -- & -- &  \\
\quad os\_world\_external & too\_thin & 20 & 0/20 & -- & -- & -- &  \\
\quad piqa\_external & too\_thin & 112 & 1/111 & -- & -- & -- &  \\
\quad scicode\_external & too\_thin & 125 & 0/125 & -- & -- & -- &  \\
\quad science\_qa\_external & too\_thin & 26 & 2/24 & -- & -- & -- &  \\
\quad simplebench\_external & too\_thin & 91 & 12/79 & -- & -- & -- &  \\
\quad spatialviz\_bench\_external & too\_thin & 8 & 8/0 & -- & -- & -- &  \\
\quad superglue\_external & too\_thin & 1 & 0/1 & -- & -- & -- &  \\
\quad swe\_bench\_verified & too\_thin & 35 & 0/35 & -- & -- & -- &  \\
\quad terminalbench\_external & too\_thin & 202 & 0/202 & -- & -- & -- &  \\
\quad the\_agent\_company\_external & too\_thin & 16 & 9/7 & -- & -- & -- &  \\
\quad trivia\_qa\_external & too\_thin & 113 & 0/113 & -- & -- & -- &  \\
\quad video\_mme\_external & too\_thin & 50 & 14/36 & -- & -- & -- &  \\
\quad vpct\_external & too\_thin & 38 & 5/33 & -- & -- & -- &  \\
\end{longtable}
}
\addtocounter{table}{-1}

\begin{figure}[htbp]
  \centering
  \includegraphics[width=\linewidth]{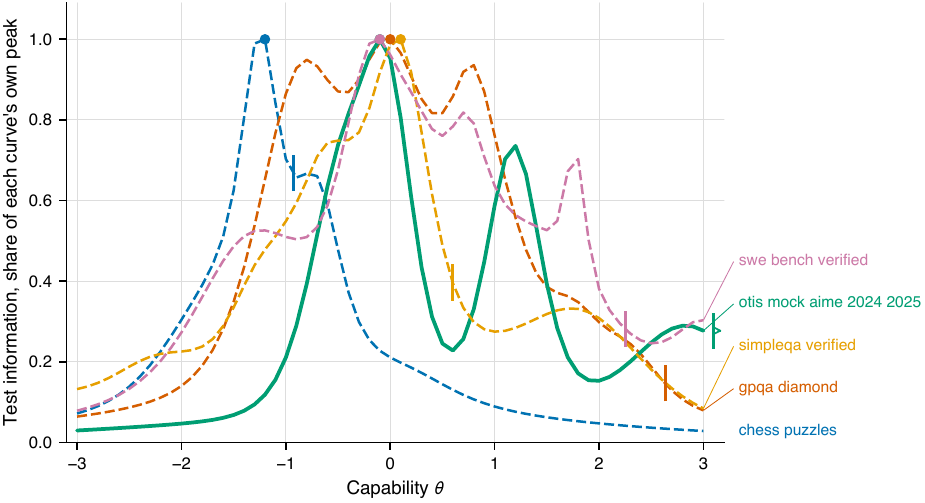}
  \caption{Item information against capability for the five benchmarks with item-level data. We normalise each curve to its own maximum, and shapes are then comparable across benchmarks whose information differs by orders of magnitude. The filled circle marks each curve's peak and the tick marks that benchmark's own ninetieth percentile of capability. Because of the normalisation the tick's height is the ratio of information there to information at the peak. Every curve peaks below its own p90, and each benchmark is least informative about exactly the models at the top of its range. The solid curve is the one benchmark the ceiling-proximity instrument dates and the dashed curves the four it does not. These curves demonstrate the mechanism without corroborating the dating. That benchmark's own p90 lies at +4.3, beyond the clipped axis. We draw its tick at the edge with an arrow pointing off the axis and carry the value in the fields.}
  \Description{Item information against capability for the five benchmarks with item-level data. We normalise each curve to its own maximum, and shapes are then comparable across benchmarks whose information differs by orders of magnitude. The filled circle marks each curve's peak and the tick marks that benchmark's own ninetieth percentile of capability. Because of the normalisation the tick's height is the ratio of information there to information at the peak. Every curve peaks below its own p90, and each benchmark is least informative about exactly the models at the top of its range. The solid curve is the one benchmark the ceiling-proximity instrument dates and the dashed curves the four it does not. These curves demonstrate the mechanism without corroborating the dating. That benchmark's own p90 lies at +4.3, beyond the clipped axis. We draw its tick at the edge with an arrow pointing off the axis and carry the value in the fields.}
  \label{fig:fa3}
\end{figure}

\begin{table}[htbp]
  \centering
  \small
  \resizebox{\ifdim\width>\linewidth\linewidth\else\width\fi}{!}{\begin{tabular}{lrrrrl}
\toprule
Benchmark & Models & Items & $\rho$ 2PL vs CRM & $\rho$ raw vs CRM & Agrees at 0.7 \\
\midrule
chess\_puzzles & 20 & 79 & 0.098 & 0.210 & no \\
gpqa\_diamond & 10 & 78 & 0.806 & 0.806 & yes \\
otis\_mock\_aime\_2024\_2025 & 10 & 11 & 0.564 & 0.807 & no \\
simpleqa\_verified & 13 & 887 & 0.731 & 0.665 & yes \\
swe\_bench\_verified & 30 & 358 & 0.728 & 0.799 & yes \\
\bottomrule
\end{tabular}
}
  \caption{Within-benchmark agreement between the item-level two-parameter fits and the continuous-response capability estimate for the five readable benchmarks. Three of five agree at rank correlation 0.7 or better, and the agreement column records the verdict. The items and raw-correlation columns attribute the two failures. One benchmark's raw accuracy tracks capability at only 0.21 within a narrow ability band, and one benchmark retains eleven usable items after we remove degenerate items.}
  \label{tab:g2-agreement}
\end{table}

\subsection{Index estimation detail}
\label{app:index-detail}

Supports Section~\ref{sec:indices}.

\begin{figure}[htbp]
  \centering
  \includegraphics[width=\linewidth]{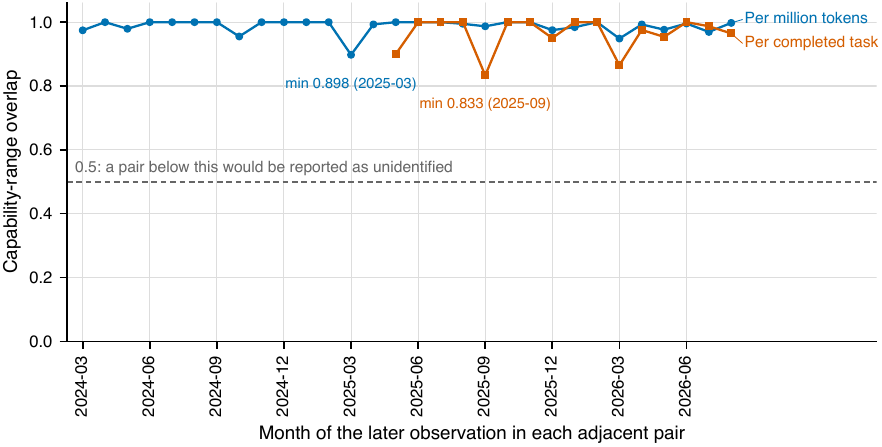}
  \caption{Capability-range overlap between adjacent months of the hedonic sample, per unit. Price variation at comparable capability identifies the month coefficients. We would report a pair below one half as unidentified. No pair approaches it. The minima are 0.898 per token and 0.833 per task, at the months annotated.}
  \Description{Capability-range overlap between adjacent months of the hedonic sample, per unit. Price variation at comparable capability identifies the month coefficients. We would report a pair below one half as unidentified. No pair approaches it. The minima are 0.898 per token and 0.833 per task, at the months annotated.}
  \label{fig:fa4}
\end{figure}

Figure~\ref{fig:fa4} shows the identification diagnostic for Section~\ref{sec:indices}. Capability-range overlap between adjacent months never falls below 0.898 per token or 0.833 per task, against an identification floor of one half.

\subsection{Reproducibility and data provenance}
\label{app:repro}

Supports Section~\ref{sec:data}.

The released bundle recomputes the panel, the capability estimate, the validity audit, every index and every figure and table from the archived provenance ledgers and page captures with one command, \texttt{make journal}, in the pinned environment. Every reported number recomputes identically, and artifacts may differ only in embedded input hashes and optimiser diagnostics. Every results artifact embeds the SHA-256 of each input it was computed from, and a rebuilt file that differs announces itself. The pinned build environment alone produces the figure PDFs. The artifacts record its library versions, and renders are byte-reproducible. The provenance ledgers, one line per observation with source URL and timestamp, ship with the dataset behind a committed checksum manifest. The trail from any published number to its source page thus survives the repository's history. Two boundaries apply. That command does not re-pull the original web pages, since the ledgers and captures ship in place of the payloads. The hand-verification and adjudication counts are review decisions, and a reader reproduces them by inspection. Tags name the frozen state, and prices are byte-identical across them. We filed the pre-registration after the freeze and before any estimation. The release links it with its complete public deviation log. A single fresh re-pull of the archival source recovers roughly 57\% of the frozen panel's rows, since the archive was sampled across many sessions. The accumulated ledgers ship in place of a re-pull for that reason.

\end{document}